\documentclass[superscriptaddress,floatfix,nofootinbib,showpacs]{revtex4-1}
\pdfoutput=1
\usepackage{amsmath,amsthm,amssymb,bm,hyperref,graphicx,color}
\usepackage[toc,page]{appendix}

\newcommand{\be}{\begin{equation}}
\newcommand{\ee}{\end{equation}}

\newcommand{\lam}{\lambda}
\newcommand{\inti}{\int_{-\infty}^{+\infty}}
\newcommand{\intp}{\int_{-\pi}^{+\pi}}
\newcommand{\gp}{\textsf{g}}
\newcommand{\6}{\partial}
\newcommand{\U}{\textsf{U}}
\newcommand{\R}{\textsf{R}}

\newcommand{\F}{\textsf{F}}

\newcommand{\ov}{\overline}
\newcommand{\la}{\lambda}

\newcommand{\st}{\kappa}

\begin{document}

\title{Correlation functions and momentum distribution of one-dimensional  hard-core anyons
in optical lattices}

\author{Ovidiu I.~P\^{a}\c{t}u}
\affiliation{Institute for Space Sciences, Bucharest-M\u{a}gurele, R
077125, Romania}
\email[E-mail:]{patu@spacescience.ro}
\pacs{05.30.Pr, 02.30.Ik, 71.10.Pm}

\begin{abstract}

We address the problem of calculating the correlation functions in a system of one-dimensional
hard-core anyons that can be experimentally realized in optical lattices. Using the summation
of form factors we have obtained Fredholm determinant representations for the time-, space-,
and temperature-dependent Green's functions which are particularly suited to numerical investigations.
In the static case we have also derived the large distance asymptotic behavior of the correlators and
computed the momentum distribution function at zero  and finite temperature. We present extensive numerical
results highlighting the  characteristic features of one-dimensional systems with fractional statistics.
\end{abstract}

\maketitle


\section{Introduction}

One-dimensional (1D) systems possess certain characteristics which makes them
extremely interesting to theoreticians and experimentalists alike. From
the theoretical point of view they are special because their simplicity
makes them amenable to exact solutions which, contrary to a naive expectation,
are characterized by  extremely rich physics. Furthermore, in recent years
numerous experimental realizations of quasi-one-dimensional  materials
opened the way for  the investigation of properties not present in the
three-dimensional world. One such remarkable feature found only in low-dimensional systems
is the presence of anyons \cite{LM,GMS,FW}, particles with statistics
interpolating between fermions and bosons. One-dimensional anyons \cite{Kundu,
BGO,BG,SSC,BGH,CM,PKA1,L2,L3,L4,SC,HZC,GHC,SC2,C, AN,HZC1,HC,WRDK,AOE1,OAE1,IT,
LMP,Gir,BGK,BCM,BFGLZ,Gr,BB,BS,MSo,YCG,SPK,Zin,RFB} represent an active area of research
with an increasing literature on the subject partly motivated by several
proposals \cite{DDL,ABVC,JBGH,KLMR,MPMP} for experiments designed to find
signatures  of fractional statistics in ultracold gases.

The model studied  in this paper \cite{AN,HZC1,HC,WRDK} can be viewed as the
generalization to arbitrary statistics of hard-core bosons on the lattice. Unlike
the continuum analog, which is represented by impenetrable Lieb-Liniger anyons
\cite{Kundu,BGO,BG,SSC,BGH,CM,PKA1,L2,L3,L4,SC,HZC,GHC,SC2,C} this model has not
been subjected to a thorough analytical investigation of
the correlation functions, especially at finite temperature.
The need for such calculations has become even more pressing in the last decade as a result of
the  plethora of experimental techniques developed  for the measurement of correlators.
Here, we fill this gap in the literature, deriving
Fredholm determinant representations for the time-, space-, and temperature-dependent
Green's functions. These representations not only represent the starting point
for the rigorous investigation of the asymptotic behavior but also can  be used
to obtain efficiently extremely accurate numerical data. We should point out that
the numerical implementation of the determinant formulas derived in this paper
can be done in less than ten lines of code in Mathematica or Maple  (see \cite{Bor})
and can be used to obtain very precise results (errors smaller than $10^{-10}$).
This level of numerical efficiency (the running time on a personal computer is of the
order of seconds) and precision cannot be achieved using other methods, such as
the ones used in \cite{HZC1,HC,WRDK}, which are also inapplicable at finite temperature.
In the case of static correlators we also derive the large distance asymptotic behavior
and compute the momentum distribution function at zero and finite temperature.
Our results show that the Green's function $\gp^{(T)}_-(m)=\langle a^\dagger_{m+1}
a_1\rangle_T$ satisfies   $\gp^{(T)}_-(m)=\ov\gp^{(T)}_-(-m)\, $ with $\mbox{Im }\gp^{(T)}_-(m)\ne 0$
for intermediate values of the statistics parameter which explains why the
groundstate momentum distribution of anyons is asymmetric \cite{HZC,SC,HZC1}
with a peak whose location in  momentum space  depends on the statistics parameter.
The location of this peak, which is present even at finite temperature, is extremely important
from the experimental point of view due to the fact that the statistics of the particles
can be inferred from its position.

The plan of the paper is as follows. In Sections \ref{Sint} and \ref{Sresults}
we present the anyonic model and the determinant representations which constitute one of the
main results of our paper. Analytical and
numerical data for the the asymptotic behaviour and momentum distribution can be
found in Section \ref{Sasy}. The derivation of the determinant representations
is being presented in Section \ref{Sdet}. We end with some conclusions in
Section \ref{Sconc}. Some technical details of the calculations
can be found in three Appendices.

\section{The model}\label{Sint}

We consider a system of one-dimensional hard-core anyons \cite{AN,HZC1,HC,WRDK} in a tight-binding
model with the Hamiltonian
\be\label{ham}
H=-J\sum_{j=1}^L\left(a_j^\dagger a_{j+1}+a_{j+1}^\dagger a_j\right)+2h \sum_{j=1}^L a_j^\dagger a_j\, ,
\ee
where $J$ is the hopping parameter, $L$ is the number of lattice sites which we will consider
to be even and $h\in[0,J)$ is the chemical potential.  The operators $a_j, a_j^\dagger$ satisfy anyonic commutation relations
\be\label{comm}
a_j a_k^\dagger=\delta_{jk}-e^{-i \pi \st \epsilon(j-k)}a_k^\dagger a_j\, ,\ \
a_j a_k=-e^{i \pi \st \epsilon(j-k)}a_k a_j\, ,\ \
a_j^\dagger a_k^\dagger=-e^{i \pi \st \epsilon(j-k)}a_k^\dagger a_j^\dagger\, ,
\ee
with
\be
 \epsilon(k)=k/|k|\, ,\ \ \ \ \epsilon(0)=0\, ,
\ee
and $\st\in[0,1]$ the statistics parameter.
For $j=k$ we have $a_j^2=(a_j^\dagger)^2=0$ (hard-core condition) and $\{a_j,a_j^\dagger\}=1$.
Varying the statistics parameter $\st$ the commutation relations (\ref{comm}) interpolate
between the ones for spinless fermions ($\st=0$) and hard-core bosons ($\st=1$). Some of the
ground state properties and relaxation dynamics of the model were studied in \cite{AN,HZC1,HC,WRDK}.
We should point out that the Hamiltonian (\ref{ham}) is a particular case  ($U=0$) of
the XXZ spin chain with fractional statistics first considered by Amico, Osterloh and Eckern
in \cite{AOE1}.

Defining the Fock  vacuum in the usual fashion
$
a_i|0\rangle=0\, ,\ \ \langle 0| a_i^\dagger=0\, ,\ \  \langle 0|0\rangle=1\, ,
$
the eigenstates of the Hamiltonian (\ref{ham}) with $N$ particles are
\be\label{eigen}
|\Psi_N(\{p\})\rangle=\frac{1}{\sqrt{N!}}\sum_{m_1=1}^L\cdots\sum_{m_N=1}^L
               \chi_N(m_1,\cdots,m_N|p_1,\cdots,p_N)a_{m_N}^\dagger\cdots a_{m_1}^\dagger|0\rangle\, ,
\ee
with $\chi_N$ the $N$-body wavefunction and $\{p\}$ the momenta of the particles.
Similar to the case of Lieb-Liniger anyons \cite{PKA1} the order in which the creation
operators appear in (\ref{eigen}) is important as we will see in the subsequent
calculation of form factors. A direct consequence of this ordering and commutation relations
(\ref{comm}) is that the exchange symmetry of the wavefunction is given by
\be\label{exchange}
\chi_N(\cdots,m_i,m_{i+1},\cdots)=-e^{i \pi \st \epsilon(m_i-m_{i+1})}\chi_N(\cdots,m_{i+1},m_{i},\cdots)\, .
\ee
The solutions of the Schr\"odinger equation
$
-J\sum_{j=1}^N[\chi_N(m_1,\cdots,m_i+1,\cdots,m_N)+\chi_N(m_1,\cdots,m_i-1,\cdots,m_N)]
  +2hN\chi_N(m_1,\cdots,m_N)=E\chi_N(m_1,\cdots,m_N)\,
$
with the appropriate symmetry (\ref{exchange}) are
\be\label{wave}
\chi_N(m_1,\cdots,m_N|\{p\})=\frac{i^{N(N-1)/2}}{\sqrt{N!}}\exp\left(i\frac{ \pi \st}{2}\sum_{1\leq a<b \leq N}\epsilon(m_a-m_b)\right)
                           \sum_{P\in S_N} (-1)^{[P]} \left(\prod_{a=1}^N e^{i m_a p_{P(a)}}\right)\, ,
\ee
with $S_N$  the group of permutations of $N$ elements and we denoted by $(-1)^{[P]}$
the signature of the permutation $P$. The $i^{N(N-1)/2}$ factor is added for convenience
so that in the bosonic limit  ($\st=1$) the wavefunctions reduce to those used
in \cite{CIKT} for hard-core bosons (XX0 spin chain). The eigenspectrum of the system is
\be
E(\{p\})=\sum_{j=1}^N \varepsilon (p_j)\, , \ \ \ \varepsilon(p)=-2J \cos p+2 h\, .
\ee

Due to the exchange symmetry (\ref{exchange}), imposing periodic boundary conditions (PBC) in
an anyonic system has nontrivial consequences as it was first noticed by Averin and
Nesteroff \cite{AN} (for a detailed discussion see  Appendix A of \cite{PKA1}). Let us
consider the simple case of two particles. If we impose PBC on the first particle
$\chi(0,m_2)=\chi(L,m_2)$ using
(\ref{exchange}) we obtain   $\chi(m_2,0)=e^{2 i \pi  \st}\chi(m_2,L)$. This shows that if
the wavefunction is periodic in the first variable then, due to the anyonic symmetry,
the wavefunction will have twisted boundary conditions in the second variable. The generalization
in the case of $N$ particles is \cite{PKA1}
\begin{align*}
\chi_N(0,m_2,\cdots,m_N)&=\chi_N(L,m_2,\cdots,m_N)\, ,\\
\chi_N(m_1,0,m_3,\cdots,m_N)&=e^{2  \pi i \st}\chi_N(m_1,L,m_3,\cdots,m_N)\, ,\\
&\ \vdots\\
\chi_N(m_1,\cdots,m_{N-1},0)&=e^{2\pi i\st (N-1)}\chi_N(m_1,\cdots,m_{N-1},L)\, .
\end{align*}
Imposing PBC on (\ref{wave}) we obtain the Bethe Ansatz Equations (BAEs) satisfied
by the momenta
\be\label{BAE}
e^{i p_a L}=e^{-i \pi \st (N-1)}\, ,\ \ a=1,\cdots,N\, .
\ee
The BAEs (\ref{BAE}) reproduce the well known equations for hard-core bosons (XX0 spin chain)
and spinless fermions for $\st=1$ and $\st=0$, respectively. Using the commutation relations (\ref{comm})
and the BAEs it is easy to show that the eigenstates (\ref{eigen}) satisfy the orthogonality
condition $\langle \Psi_{N_1}|\Psi_{N_2}\rangle=0$ if $N_1\ne N_2$ and $\langle \Psi_{N}(\{p\})|\Psi_{N}(\{p'\})\rangle=0$
if $\{p\}\ne\{p'\}$. The normalization is
 \footnote{ More precisely if $\{p'\}=R\{p\}$ where $R\in S_N$ we have
$\langle \Psi_{N}(\{p\})|\Psi_{N}(\{p'\})\rangle=L^N (-1)^R\, .$}
\be
\langle \Psi_{N}(\{p\})|\Psi_{N}(\{p\})\rangle=L^N\, .
\ee

Introducing the notation
$
\{[x]\}=\gamma\, , \mbox{ if }  x=2\pi\times integer +2\pi \gamma\, ,\ \gamma\in(-1/2,1/2]\, ,
$
the values of the momenta in the groundstate with $N$ particles, $N$ odd,  are
\be\label{gs}
p_a=\frac{2\pi}{L}\left(-\frac{N+1}{2}+j\right)+\frac{2\pi \delta}{L}\, ,\ \ \delta=\{[-\pi\st(N-1)]\}\, , \ \  \ j=1,\cdots, N\, .
\ee
In the thermodynamic limit the momenta fill densely the Fermi zone $[-k_F,k_F]$ with $k_F=\arccos(h/J)$ the
Fermi momentum.  We remind the reader that $h\in[0,J)$ and the system is gapless.
For $h>J$ the structure of the ground-state is different and will not be considered in this paper.
For  $h\in[0,J)$ the thermodynamic behavior of hard-core anyons and
spinless fermions is the same.


\section{Determinant representation for the correlation functions}\label{Sresults}

We are interested in calculating the thermodynamic limit of the time-, space-, and temperature-
dependent Green's functions
\begin{subequations}\label{defg}
\begin{align}
\gp_+^{(T)}(m,t)&\equiv
\frac{ \textsf{tr }\left[e^{-H/T}a_{m_2}(t_2) a_{m_1}^\dagger(t_1) \right]}{\textsf{tr }[e^{-H/T}]}\, ,\\
\gp_-^{(T)}(m,t)&\equiv
\frac{ \textsf{tr }\left[e^{-H/T}a_{m_2}^\dagger(t_2) a_{m_1}(t_1) \right]}{\textsf{tr }[e^{-H/T}]}\, ,
\end{align}
\end{subequations}
where  $a_m(t)=e^{i H t} a_m e^{-i H t}$, $a_m^\dagger(t)=e^{i H t} a_m^\dagger e^{-i H t}$ and
$m=m_2-m_1\, ,\  t=t_2-t_1$.  Using the
summation of form-factors we were able to express these correlators as Fredholm determinants
which, as we will show in Section \ref{Sasy}, can be used to obtain extremely precise numerical data.
 We would like to stress the fact that our result is exact and does not employ any
approximations. The derivation of the determinant representation is quite involved, therefore,
in this section, we are going to present only the final results. The interested reader can find
the full derivation in Section \ref{Sdet}.

The correlation function $\gp_+^{(T)}(m,t)$ has the following representation
in terms of Fredholm determinants (for the definition of a Fredholm determinant
see Section \ref{Sther}):
\begin{align}\label{gplus}
\gp_+^{(T)}(m,t)&=e^{-2 i h t} \left[G(m,t)+\frac{\6}{\6 z}\right]
\left.\det\left(1+\hat{\textsf{V}}_T-z\hat\R_T^{(+)}\right)\right|_{z=0}\, ,\nonumber\\
&=e^{-2 i h t} \left[(G(m,t)-1)\det(1+\hat{\textsf{V}}_T)+\det(1+\hat{\textsf{V}}_T-\hat\R_T^{(+)})\right]\, ,
\end{align}
with $\hat{\textsf{V}}_T$ and $\hat\R_T^{(+)}$  integral operators acting  on the
interval $[-\pi,\pi]$
\be\label{action}
\left(\hat{\textsf{V}}_T f\right)(p)=\frac{1}{2\pi}\intp \textsf{V}_T(p,p')f(p') dp'\, ,\ \ \
\left(\hat\R_T^{(+)}f\right)(p)=\frac{1}{2\pi}\intp \R^{(+)}_T(p,p')f(p') dp'\, ,
\ee
with kernels
\begin{subequations}\label{kernelsp}
\begin{align}
\textsf{V}_T(p,p')&=\sin^2\left(\frac{\pi\st}{2}\right)
\left[\frac{E_+^T(p)E_-^T(p')-E_-^T(p)E_+^T(p')}{\tan\frac{1}{2}(p-p')}-G(m,t)E_-^T(p)E_-^T(p')\right]\, ,\label{vt}\\
\R_T^{(+)}(p,p')&=\sin^2\left(\frac{\pi\st}{2}\right)E_+^T(p)E_+^T(p')\, .
\end{align}
\end{subequations}
The functions appearing in (\ref{kernelsp}) are defined as
\begin{align}
G(m,t)&=\frac{1}{2\pi}\intp dp\, \ e^{i m p +2i J t\cos p}\\
E_-^T(p)\equiv E_-^T(p,m,t)&=\sqrt{\theta(p)}\, \,e^{-i m p/2-i J t \cos p}\, ,\\
E_+^T(p)\equiv E_+^T(p,m,t)&=E^T(p)\, E_-^T(p)\, ,\\
E^T(p)\equiv E^T(p,m,t)&=\mbox{P.V.}\frac{1}{2\pi}\intp \frac{e^{i m q +2i J t\cos q}}{\tan\frac{1}{2}(q-p)}\, dq
 -\cot\left(\frac{\pi\st}{2}\right)e^{i m p +2i J t\cos p}\, .
\end{align}
with $\mbox{P.V.}$ denoting the principal value of the integral and
$\theta(p)\equiv \theta(p,h,T)$ is the Fermi function
\be\label{fermi}
\theta(p)=\frac{1}{1+e^{\frac{-2J \cos p+2h}{T}}}\, .
\ee

A similar representation is obtained for the $\gp_-^{(T)}(m,t)$ correlation function
\begin{align}\label{gminus}
\gp_-^{(T)}(m,t)&=e^{2 i h t}\frac{\6}{\6 z}
\left.\det\left(1+\hat{\textsf{V}}_T+z\hat\R_T^{(-)}\right)\right|_{z=0}\, ,\nonumber\\
&=e^{2 i h t} \left[\det(1+\hat{\textsf{V}}_T+\hat\R_T^{(-)})-\det(1+\hat{\textsf{V}}_T)\right]\, .
\end{align}
In (\ref{gminus}), $\hat{\textsf{V}}_T$ is the same integral operator which appears in
the determinant representation of $\gp_+^{(T)}(m,t)$.  $\hat\R_T^{(-)}$ is an integral
operators which acts on  $[-\pi,\pi]$
\[
\left(\hat\R_T^{(-)}f\right)(p)=\frac{1}{2\pi}\intp \R^{(-)}_T(p,p')f(p') dp'\, ,
\]
with kernel
\be
\R_T^{(-)}(p,p')=E_-^T(p)E_-^T(p')\, .
\ee
The main difference in this case is that the $E_+^T(p)$ function which enters the definition of the kernel (\ref{vt}) is now
defined as (note the sign change)
\be
E^T(p)\equiv E^T(p,m,t)=\mbox{P.V.}\frac{1}{2\pi}\intp \frac{e^{i m q +2i J t\cos q}}{\tan\frac{1}{2}(q-p)}\, dq
 +\cot\left(\frac{\pi\st}{2}\right)e^{i m p +2i J t\cos p}\, .
\ee
In the $\st\rightarrow 1$ limit (\ref{gplus}) and  (\ref{gminus}) reproduce the well-known
results for hard-core bosons  \cite{CIKT} (note that in our notation,
$\gp_+^{(T)}(m,t)$ and $\gp_-^{(T)}(m,t)$ correspond to $\langle\sigma_m^+(t)\sigma_0^-(0)\rangle_T$
and  $\langle\sigma_m^-(t)\sigma_0^+(0)\rangle_T$ of \cite{CIKT}). At $\st=0$ we obtain the
results for spinless free fermions on the lattice.

\subsection{Static limit}

Certain simplifications occur in the static limit $t=0$. Due to the relation
\be
\gp_+^{(T)}(m,0)=\delta_{m,0}-e^{-i\pi\st\epsilon(m)} \gp_-^{(T)}(-m,0)\, ,
\ee
it will be sufficient to consider only $\gp_-^{(T)}(m,0)$. In this limit we have
$G(m,0)=\delta_{m,0}\, ,$ $E_-^T(p,m,0)=\sqrt{\theta(p)} e^{-im p/2}\, ,$ and
\[
E_+^T(p,m,0)=\left\{
\begin{array}{ll}
i\sqrt{\theta(p)}(1-i\cot(\pi\st/2))e^{i m p/2}\, , & m>0\, ,\\
-\sqrt{\theta(p)}\cot(\pi\st/2)\, ,                 & m=0\, ,\\
-i\sqrt{\theta(p)}(1+i\cot(\pi\st/2))e^{-i |m|p/2}\, , & m<0\, .\\
\end{array}
\right.
\]
The static limit of (\ref{gplus}) for $m>0$ is
\be\label{gplusstatic}
\gp_-^{(T)}(m,0)=\det(1+\hat{\textsf{v}}_T+\hat{\textsf{r}}^{(-)}_T)-\det(1+\hat{\textsf{v}}_T)\, ,
\ee
with $\hat{\textsf{v}}_T$ and $\hat{\textsf{r}}^{(-)}_T$ integral operators acting on
$[-\pi,\pi]$ like in Eq.~(\ref{action}) and kernels
\begin{align}
\textsf{v}_T(p,p')&=(e^{i\pi\st}-1)\sqrt{\theta(p)}\, \frac{\sin\frac{1}{2}m(p-p')}
{\tan\frac{1}{2}(p-p')}\, \sqrt{\theta(p')}\, ,\nonumber\\
\textsf{r}^{(-)}_T(p,p')&=\sqrt{\theta(p)}\, e^{-\frac{i m}{2} (p+p')}\, \sqrt{\theta(p')}\, .
\end{align}
In the case of $m<0$ the representation (\ref{gplusstatic}) remains valid but now the kernels
of the integral operators are
\begin{align}
\textsf{v}_T(p,p')&=(e^{-i\pi\st}-1)\sqrt{\theta(p)}\, \frac{\sin\frac{1}{2}|m|(p-p')}
{\tan\frac{1}{2}(p-p')}\, \sqrt{\theta(p')}\, ,\nonumber\\
\textsf{r}^{(-)}_T(p,p')&=\sqrt{\theta(p)}\, e^{\frac{i |m|}{2} (p+p')}\, \sqrt{\theta(p')}\, .
\end{align}
This shows that for intermediate statistics $\mbox{Im } \gp_-^{(T)}(m,0)\ne 0$  and
\be\label{cc}
\gp_-^{(T)}(m,0)=\ov\gp_-^{(T)}(-m,0)\, ,
\ee
with the bar denoting complex conjugation. As we will see in Section \ref{Smom} the direct consequence of
this relation is that the  momentum distribution of anyons is no longer symmetric around the origin.

\subsection{Zero temperature limit}

The zero temperature limit results can be derived easily noticing that at $T=0$ the
Fermi function (\ref{fermi}) becomes the characteristic function of the interval $[-k_F,k_F]$
with $k_F=\arccos(h/J)$. Therefore, the groundstate Green's functions have the same determinant
representations (\ref{gplus}),(\ref{gminus}), (\ref{gplusstatic})  with  integral operators
acting on $[-k_F,k_F]$
\[
\left(\hat{\textsf{V}} f\right)(p)=\frac{1}{2\pi}\int_{-k_F}^{k_F} \textsf{V}(p,p')f(p') dp'\, ,\ \ \
\left(\hat{\R}^{(\pm)} f\right)(p)=\frac{1}{2\pi}\int_{-k_F}^{k_F} \R^{(\pm)}(p,p')f(p') dp'\, ,
\]
and similar expressions for $\hat{\textsf{v}}\, , \hat{\textsf{r}}^{(+)}$ and
$E_-^T(p)\rightarrow E_-(p)=e^{-i m p/2-i J t \cos p}$.


\section{Asymptotic behavior of static correlators}\label{Sasy}

Similar Fredholm determinant representations for the correlation functions of other integrable
systems were previously obtained in the  case of impenetrable bosons \cite{Lenard,IIK1},
XX0 spin chain \cite{CIKT}, two-component bosons and fermions \cite{IP} and Lieb-Liniger
anyons \cite{L2}. In all these cases, including ours, the relevant  integral operators
belong to the so-called ``integrable" class of integral operators \cite{IIKS2,KBI,HI} and
as a result the correlation functions satisfy a completely integrable classical system of
differential equations.  The determinant representation and the associated differential
equations represent the basis for the rigorous investigation of the large time and distance
asymptotic behavior of the correlators via the solution of a certain matrix Riemann-Hilbert
problem. This rather involved program was implemented for the impenetrable Bose gas in
\cite{IIK1,IIKV,KBI}, for the XX0 spin chain in \cite{IIKS}, for the two-component fermions
in \cite{BL,GIKP,CZ} and for Lieb-Linger anyons in \cite{L3,L4}.

It would be naturally to expect a comparable number of papers in the literature
devoted to the numerical exploration of the determinant representations which
would allow for a better understanding of the intermediate distance regime
which is inaccessible by analytical methods. This is unfortunately not true (two notable
exceptions being \cite{CSZ,Z}). While this might be attributed to the possible
uncontrollable  errors in the evaluation of an infinite determinant, recently
Bornemann \cite{Bor} provided a simple and easily implementable method which
allows for very precise numerical  evaluations of such representations. This method,
which is based on the Nystr\"om solution of the Fredholm integral equations of the second kind
with the Gauss-Legendre as the quadrature rule, will be used to calculate
the short and intermediate distance static correlation functions and also to check
the validity of the large distance asymptotics which we will derive below.

\subsection{Asymptotic behavior at zero temperature}

At zero temperature we expect the system to be critical and the large
distance asymptotic behavior of the correlation functions can be derived
using Conformal Field Theory ideas. More precisely we are going to use
Cardy's result \cite{Cardy}  relating the conformal dimensions of the
conformal fields present in the theory from the finite size corrections
of the low-lying excitations of the system (a detailed presentation of the
method can be found in  Chap. XVIII of \cite{KBI}).
\begin{figure}[h]
\includegraphics[width=\linewidth,height=18cm]{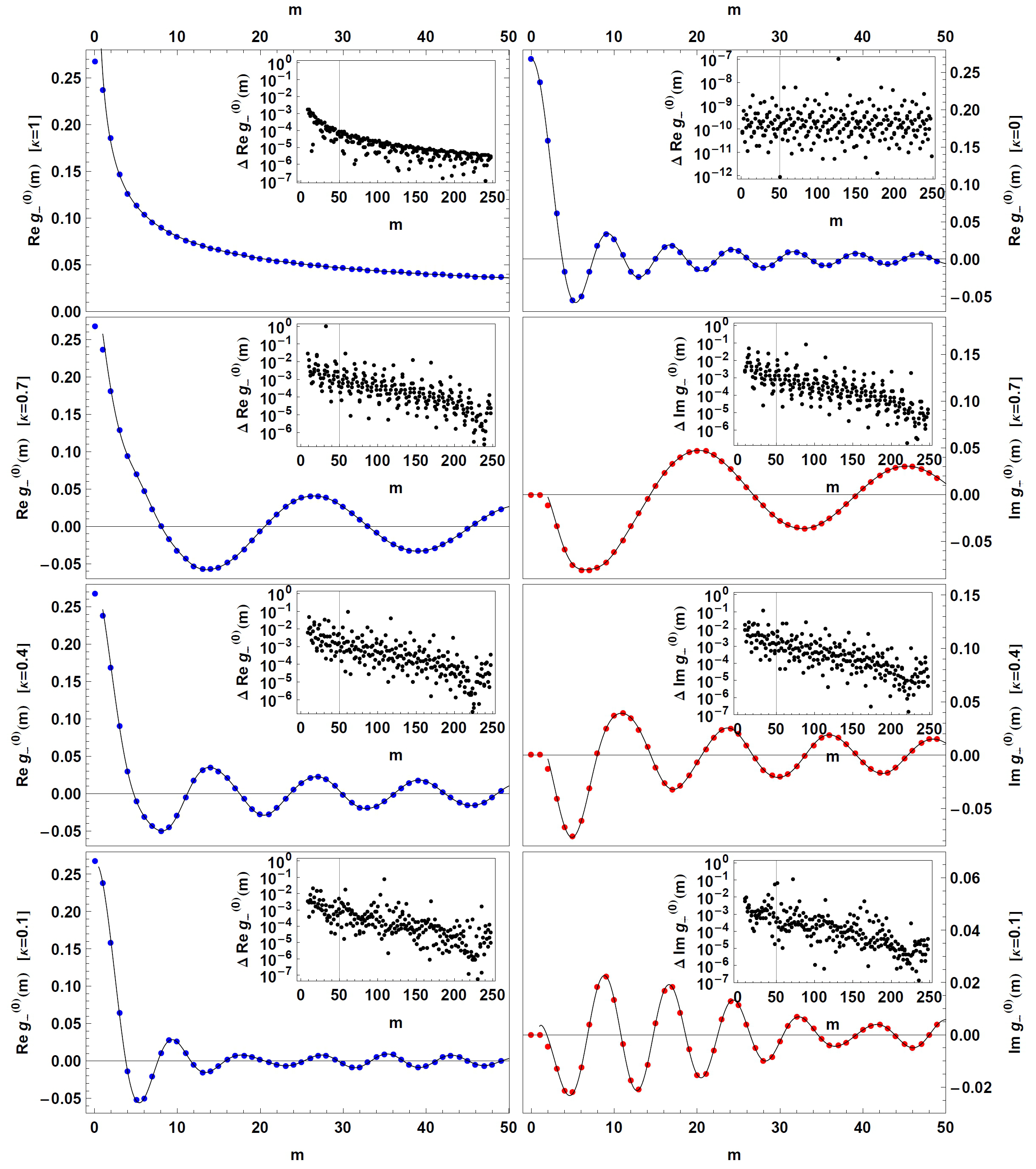}
\caption{(Color online) Plot of the real and imaginary part of the zero temperature correlation
function $\gp_-^{(0)}(m)$ (blue and red dots) and  asymptotic behavior (thin line)
given by Eq.~\ref{t0a}  for $h=4/3,\, J=2$ and different values of the statistics parameter.
The results for hard-core bosons ($\st=1$) and spinless free fermions ($\st=0$) are
presented in the top panels.
The insets contain the relative errors of the asymptotic formula for $m$ up to 250
(the errors of the data presented in the main panels are shown up to the vertical line at $m=50$).
(Distance $m$ in units of $a_0$ the lattice constant which,
for convenience, is set to 1.)}
\label{T0}
\end{figure}
For our system we have three type of low-lying excitations: addition of
one or more  particles into the system with quantum number $\Delta N$, backscattering
of $d$  particles over the Fermi ''sea" with quantum number $d$ and
''particle-hole" excitations close to the left or right  Fermi surface
with quantum numbers $N^\pm$. The asymptotic behavior of the correlation
functions is then
\[
\langle \phi(m)\phi(0)\rangle=\sum_{Q} A(Q)\frac{ e^{i p_Q m}}{|m|^{2\Delta_Q^++2\Delta_Q^-}}\, ,
\]
where $Q=\{\Delta N,d,N^\pm\}$,  $A(Q)$ are amplitudes which cannot be obtained using
this method, $p_Q$ is the macroscopic part of the momentum
gap and the conformal dimensions $\Delta_Q^\pm$ can be obtained from the finite
size corrections of the energy and momentum using ($v_F$ is the Fermi velocity)
\begin{align}
P_Q-P_0&=p_Q+\frac{2\pi}{L}\left(\Delta_Q^+-\Delta_Q^-\right)\, ,\nonumber\\
E_Q-E_0&=\frac{2\pi v_F}{L}\left(\Delta_Q^++\Delta_Q^-\right)\, .
\end{align}
\begin{figure}[h]
\includegraphics[width=\linewidth]{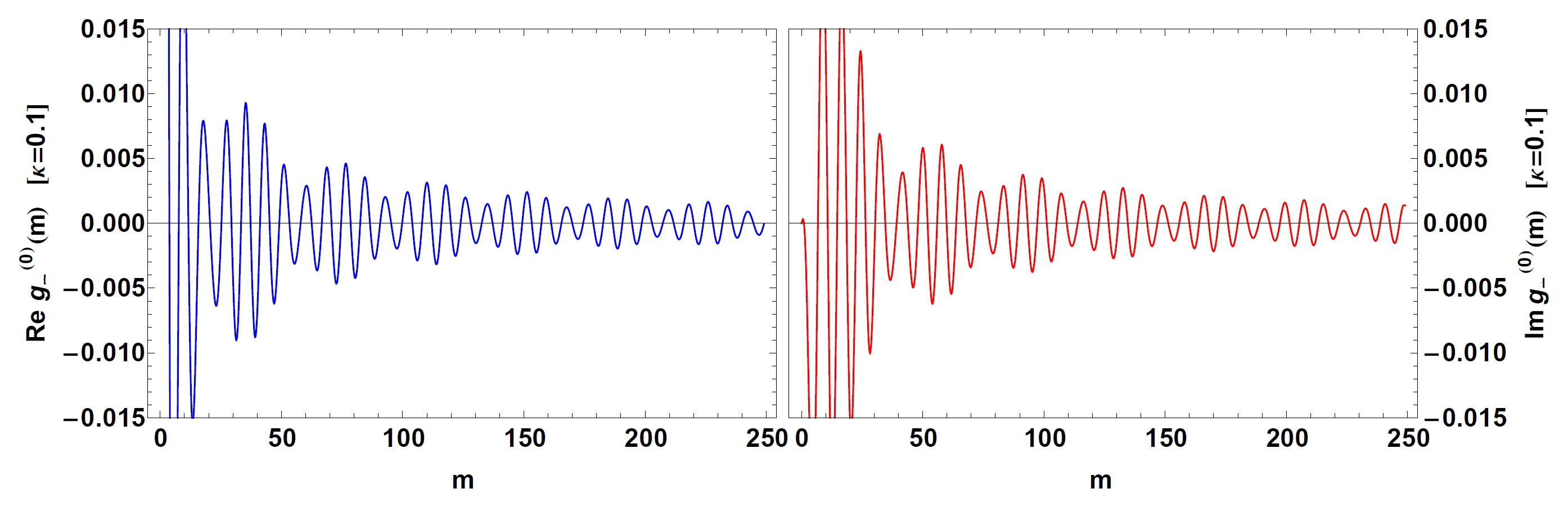}
\caption{Real and imaginary part of the correlation function $\gp_-^{(0)}(m)$
for $h=4/3,\, J=2$ and  $\st=0.1$ showing fermionic beats. (Distance $m$ in units of $a_0$.)}\label{T0b}
\end{figure}
The derivation of the finite size corrections in our model is very similar with the
one performed for the XX0 spin chain \footnote{Ref. \cite{KBI} considers the more general
case of the XXZ spin chain. The results relevant for us are obtained considering
$\Delta=0$ and the dressed charge $\mathcal{Z}=1$.} (Chap. II of \cite{KBI}) and can be found in
Appendix \ref{Afsc}.  The central charge of the model is  equal to one and
\begin{subequations}\label{corrections}
\begin{align}
P_{\Delta N=1,d,N^\pm}-P_0&=2k_F\left(d+\frac{1-\st}{2}\right)+\frac{2\pi}{L}\left[\left(d+\frac{1-\st}{2}\right)+N^++N^-\right]\, ,\nonumber\\
E_{\Delta N=1,d,N^\pm}-E_0&=\frac{2\pi v_F}{L}\left[\left(\frac{1}{2}\right)^2+\left(d+\frac{1-\st}{2}\right)^2+N^++N^-\right]\, .
\end{align}
\end{subequations}
Neglecting the contributions coming from the $N^\pm$ terms we obtain the following asymptotic expansion
for the Green's function
\be\label{fe}
\gp_-^{(0)}(m)\sim \sum_{d\in\mathbb{Z}}\, A(d) \frac{e^{i k_F[2d +(1-\st)] m}}{|m|^{\frac{[2d+(1-\st)]^2}{2}+\frac{1}{2}}}\, ,
\ee
 where $d$ is the number of backscattered particles.
For $\st=1$ this expansion reduces to the Green's function asymptotic expansion
of hard-core bosons with leading term $\gp_-^{(0)}(m)\sim 1/|m|^{1/2}.$ At the free-fermionic point,
$\st =0$, neglecting all the terms except $d=0,-1$ we get $\gp_-^{(0)}(m)\sim \sin(k_F m)/m$ which is
in fact the exact result modulo the amplitude.
For intermediate values of the statistics parameter
the leading term of the expansion, given by $d=0$, is oscillating with a wavevector  proportional
to $1-\st$. This is a general characteristic of 1D anyonic systems first observed by Calabrese
and Mintchev \cite{CM} and it can be seen in Fig. \ref{T0}. The asymptotic behavior plotted is
\be\label{t0a}
\tilde\gp_-^{(0)}(m)=  a\,  \frac{e^{i[(k_F(1-\st) m-b}]}{|m|^{\frac{(1-\st)^2}{2}+\frac{1}{2}}}+
c\, \frac{e^{i [k_F(-2 +(1-\st)) m+d]}}{|m|^{\frac{[-2+(1-\st)]^2}{2}+\frac{1}{2}}}\, ,
\ee
with $a,b,c,d$ real parameters obtained using a fitting procedure and the relative errors defined
in the usual fashion $\Delta\mbox{Re } \gp_-^{(0)}(m)=| \mbox{Re } \gp_-^{(0)}(m)-\mbox{Re }
\tilde\gp_-^{(0)}(m)| / | \mbox{Re } \gp_-^{(0)}(m)|$ and a  similar expression for the imaginary
part. The correlation function $\gp_-^{(0)}(m)$ was computed from the zero temperature limit
of Eq.~(\ref{gplusstatic}). Using the method presented in \cite{Bor} for the numerical implementation
of Fredholm determinants  we were able to obtain extremely accurate values (absolute errors smaller than $10^{-10}$).

Another interesting feature specific to correlation functions of 1D anyonic systems (see \cite{CM})
is the presence of fermionic beats for values of the statistics parameter close to $0$.
In this region we can see from the r.h.s. of (\ref{t0a}) that we have two oscillations
with almost equal amplitudes and wavevectors producing  beating effects. This phenomenon can
be seen in  Fig.~\ref{T0b} for $\st=0.1$.

\subsection{Asymptotic behavior at finite temperature}

\begin{figure}[t]
\includegraphics[width=\linewidth,height=18cm]{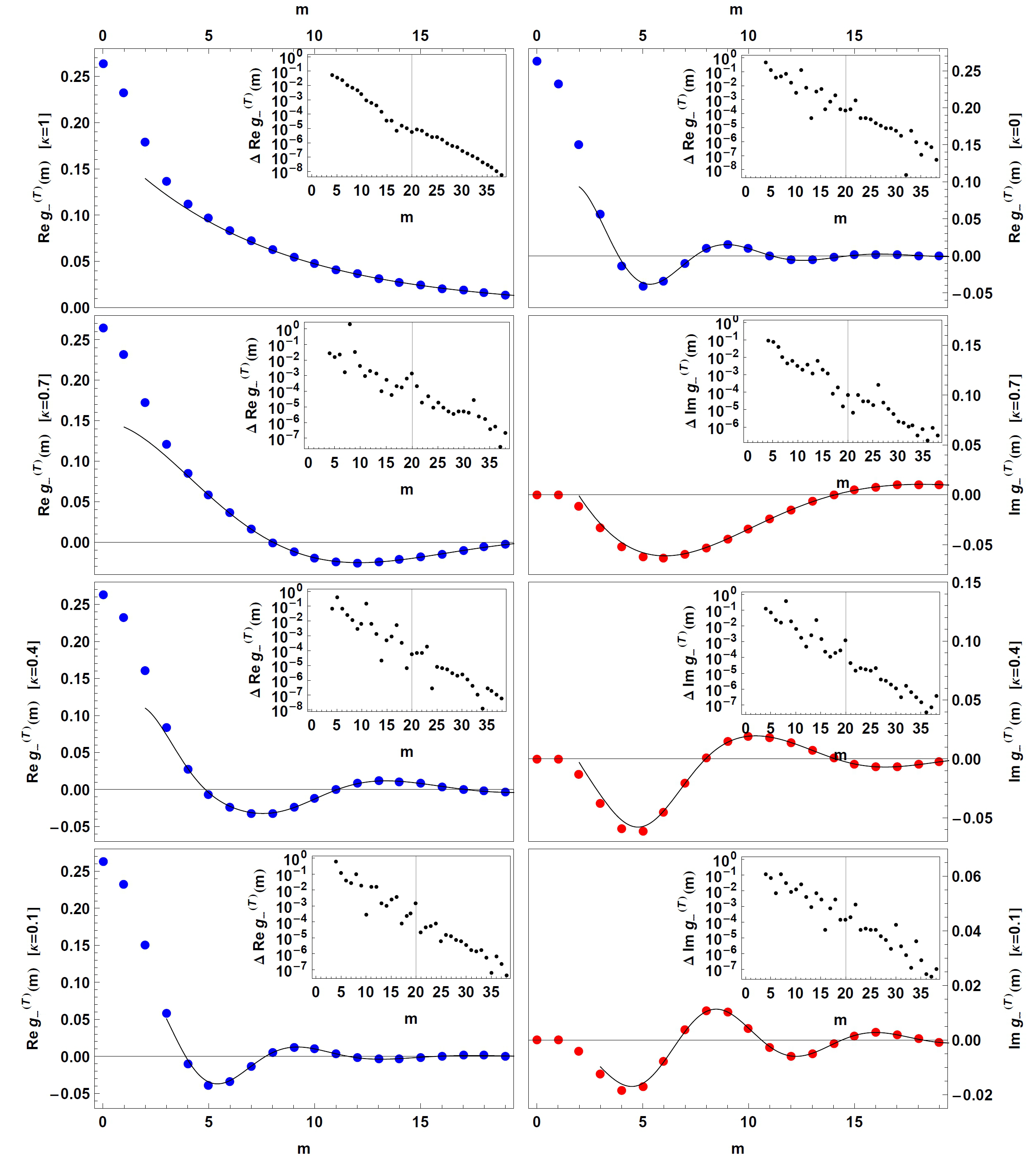}
\caption{(Color online) Plot of the real and imaginary part of the  temperature dependent
correlation function $\gp_-^{(T)}(m)$ (blue and red dots) and  asymptotic behavior (thin line)
given by Eq.~\ref{ta} for $h=4/3\, , J=2\, $ temperature $T=1/4$ and different values of the statistics parameter.
The results for hard-core bosons ($\st=1$) and spinless free fermions ($\st=0$) are
presented in the top panels.
The insets contain the relative errors of the asymptotic formula for $m$ up to 40
(the errors of the data presented in the main panels are shown up to the vertical line at $m=20$).
The insets contain the relative errors of the asymptotic formula. (Distance $m$ in units of $a_0$.)}
\label{T025}
\end{figure}

At low-temperatures the asymptotic behavior of $\gp_-^{(T)}(m)$
can  be derived from the zero temperature result (\ref{fe}) by replacing $1/|m|$ with $1/\sinh(\pi v_F m/T)$
with $v_F$ the Fermi velocity. However, at higher temperatures the CFT description is no longer valid and
we need to use a different method.
Here, we present an heuristic derivation of the temperature dependent asymptotic
expansion of the Green's function  which is based on the similar results obtained rigorously for
impenetrable Lieb-Liniger anyons \cite{L3}. Even though the considerations below
are nothing more than an educated guess, the numerical data presented in Fig.~\ref{T025} show
that our result, Eq.~(\ref{ta}), is nevertheless correct.
In the case of Lieb-Liniger impenetrable anyons, the main term responsible for the exponential decay
of the correlators, denoted by $C(\beta,\st)$ in Eq.~3 of \cite{L3}, contained the logarithm of the ratio between the Fermi distribution
and an ``anyonic" distribution function  interpolating between the Fermi and Bose
distributions. We will assume that for our model the lattice equivalent of this anyonic distribution function
is  given by
\be
\theta(p,\st)=\frac{1}{e^{\frac{-2J \cos p+2h}{T}}-e^{i\pi(1-\st)}}\, ,
\ee
in terms of which we can define the lattice analog of $C(\beta,\st)$
\be
C(\st)=\frac{1}{2\pi}\intp \log\left(\frac{\theta(p,\st)}{\theta(p,0)}\right) dp\, .
\ee
An important role is played by the zeroes of $\theta(p,\st)^{-1}$ closest  to the real axis
and situated in the  upper half plane denoted by

\begin{align}
\la_0(\st)&=\arccos\left[\frac{1}{2J}(2h -i(1-\st)\pi T)\right]\, ,\nonumber\\
\la_{-1}(\st)&=-\arccos\left[\frac{1}{2J}(2h +i(2\pi T-(1-\st)\pi T))\right]\, .
\end{align}
Then, the asymptotic behavior of the correlation function obtained in direct
analogy with the continuum result \cite{L3} is given by
\footnote{ There is a typo in Eq.~(6) of \cite{L3}. The correct version is
$
\lam_j=(-1)^j\left(\beta+\sqrt{\beta^2+\pi^2[\kappa+2j]^2}\right)^{1/2}/
\sqrt{2} +\cdots .
$
}
\be\label{ta}
\tilde\gp_-^{(T)}(m)\simeq (a+ib)\, e^{-m[C(\st)-i\la_0(\st)]}+(c+id)\, e^{-m[C(\st)-i\la_{-1}(\st)]}\, ,\ m>0\, ,
\ee
with $a,b,c,d$ real parameters.
For $\st=1$ the second term in the r.h.s. of Eq.~(\ref{ta})
is much smaller than the first term and we obtain the known result for hard-core bosons
$
\tilde\gp_-^{(T)}(m,\st=1)\simeq a\, e^{\frac{m}{2\pi}\intp\log\left|\tanh\left(\frac{h-J\cos p}{T}\right)\right|\, dp}\,
$
\cite{IIKS}. In the fermionic limit $C(\st=0)=0$ and we have
$ \tilde\gp_-^{(T)}(m,\st=0)\simeq a\, e^{ i m\la_0(0)}+c\, e^{ i m\la_{-1}(0)}\, $
with $\la_j= (-1)^j \arccos\left[\frac{1}{2J}(2h -(-1)^j i\pi T)\right]\, , \ j=0,-1$.
This result reproduces the first two terms of the asymptotic expansion of free fermions
which can also be derived  starting from  $ \gp_-^{(T)}(m,\st=0)\simeq \intp e^{i p m }\theta(p)\,  dp$
and moving the integration contour in the upper half-plane for $m>0$.
Plots of the correlation function $\gp_-^{(T)}(m)$   and the asymptotic expansion
Eq.~(\ref{ta}) for $\st=0,1$ are shown in the top panels of Fig.~\ref{T025}.
Additional numerical checks
of our  asymptotic expansion for intermediate values of the statistics parameter
are presented in the lower panels of Fig.~\ref{T025}. For values of $\st$ in the interval $[0.4,1]$ it is
sufficient to fit the data using only the first term in the r.h.s. of (\ref{ta})
obtaining extremely accurate results (relative errors of $10^{-8}$ for $m=40$ and $\st=0.4$).
As $\st$ approaches the fermionic point, $\st=0$  we need to use both terms in order
to obtain accurate results (relative errors of $10^{-8}$ for $m=40$ and $\st=0.1$).
The oscillatory behavior of the correlation functions is still present
but at finite temperature the damping is exponential compared to the algebraic decay
at zero temperature. The rate of exponential decay is a decreasing function of $\st$
being maximum for spinless fermions.

\subsection{Momentum distribution at zero and finite temperature}\label{Smom}

The accurate knowledge of the large distance asymptotic behavior allows us to compute the momentum
distribution function defined as
\be
n(k)=\frac{1}{2\pi}\sum_{m=-\infty}^\infty e^{i k m}\,  \gp_{-}(m)\, .
\ee
A direct consequence of Eq.~(\ref{cc}) is that $\mbox{Re } \gp_{-}(-m)=\mbox{Re } \gp_{-}(m)$ and
$\mbox{Im } \gp_{-}(-m)=-\mbox{Im } \gp_{-}(m)$ which means that
\be\label{asym}
n(k)=\frac{1}{\pi}\left[\frac{\mbox{Re } \gp_{-}(0)}{2}+\sum_{m=1}^\infty \mbox{Re } \gp_{-}(m)\cos(k m)
-\sum_{m=1}^\infty \mbox{Im } \gp_{-}(m)\sin(k m)\right]\, .
\ee
\begin{figure}
\includegraphics[width=\linewidth,height=14cm]{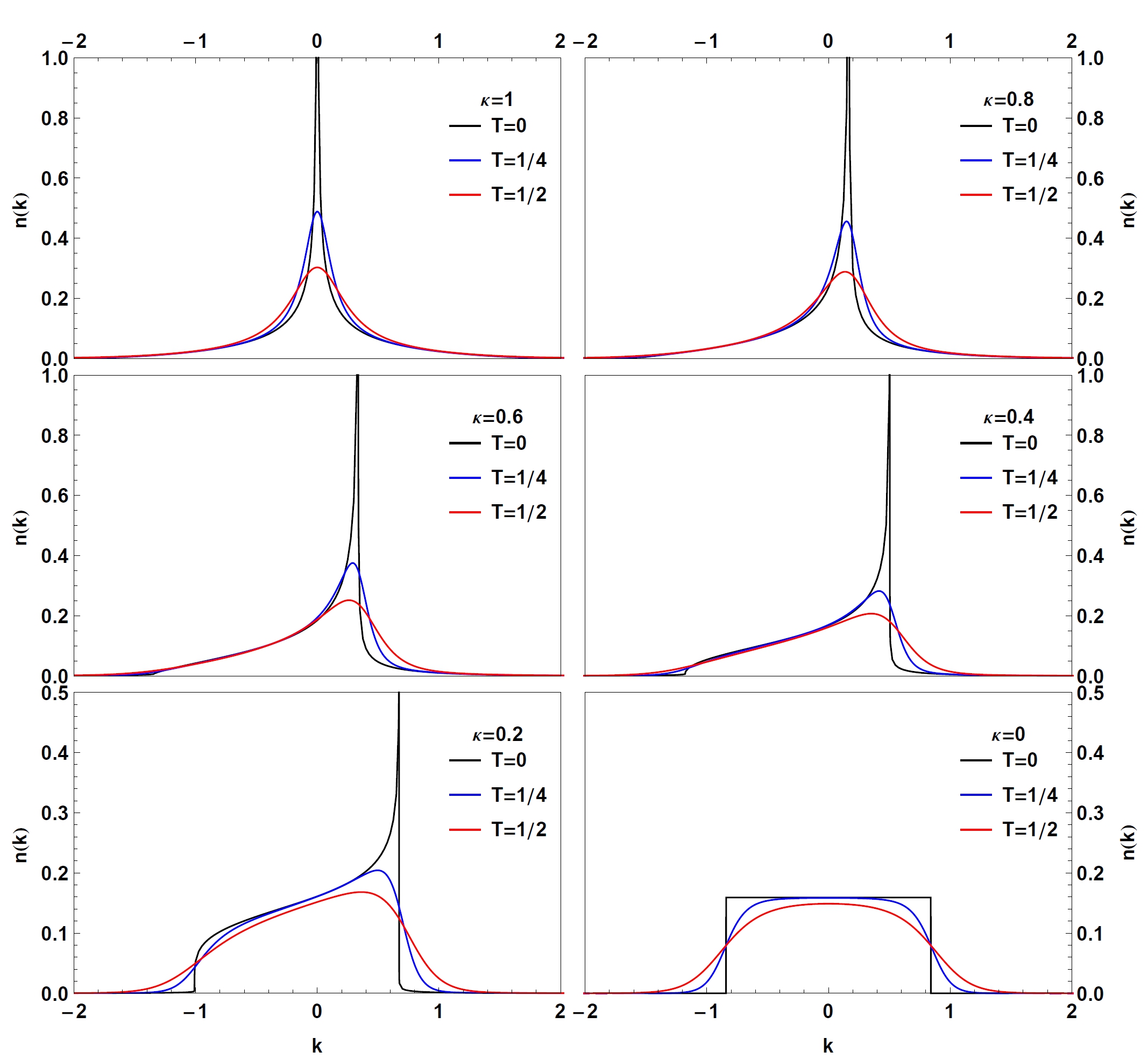}
\caption{(Color online) Momentum distribution functions  for $h=4/3\, , J=2\, $ temperature
$T=0\, ,1/4\, ,1/2$ and different values of the statistics parameter. (Momentum $k$ in units of $1/a_0$.)
}
\label{mom}
\end{figure}
For hard-core bosons and spinless fermions $\mbox{Im } \gp_{-}(m)=0$ and, therefore, the
momentum distribution is symmetric with respect to the $k$ axis. However, in the case
of anyonic systems $\mbox{Im } \gp_{-}(m)\ne 0$ and the momentum distribution is asymmetric
(see also \cite{SC,HZC,GHC}). Also, from the
asymptotic behavior (\ref{t0a}) we expect that at zero temperature the momentum distribution
will have a singularity at $k=(1-\st)k_F$ and a weaker singularity at $[-2+(1-\st)]k_F$.
The numerical results presented in Fig. \ref{mom} confirm these theoretical predictions.
At zero temperature we can see clearly how the  peak present at  $k=0$
for $\st=1$ (hard-core bosons) decreases and  moves to the right as $(1-\st)k_F$  and
becomes the discontinuity at $k_F$ of the momentum distributions for spinless fermions.
The weak singularity at $-2k_F$ for $\st=1$ manifests itself in the derivative of $n(k)$
which becomes sharper with decreasing $\st$ and becomes the discontinuity of the momentum
distribution at $-k_F$ (see also \cite{SC}). At finite temperature the momentum distribution
gets smoother and wider. Even so, the peaks at $(1-\st)k_F$ still remain visible and can
be experimentally detected.


\section{Derivation of the determinant representation for the correlation functions}\label{Sdet}

The determinant representations presented in Section \ref{Sresults} were derived using the
method known as the summation of form-factors \cite{IIK1,IP,L2,KBI} which is briefly sketched below
(an alternative method based  on Wick's theorem can be found in \cite{ZCG}).
We start from the finite lattice expression of the Green's function (we focus on $\gp_+^{(T)}(m,t)$,
the $\gp_-^{(T)}(m,t)$ case can be treated along similar lines)
\be\label{i1}
\gp_+^{(T)}(m,t)=\frac{1}{\textsf{tr }[e^{-H/T}]}\sum_{N=0}^L\sum_{p_1<\cdots<p_N}e^{-\frac{E_N(\{p\})}{T}}
\langle a_{m_2}(t_2) a_{m_1}^\dagger(t_1)\rangle _N
\, ,
\ee
where we have used the short notation  $\langle a_{m_2}(t_2) a_{m_1}^\dagger(t_1)\rangle _N=
\langle\Psi_N(\{p\})|a_{m_2}(t_2) a_{m_1}^\dagger(t_1)|\Psi_N(\{p\})\rangle/\langle\Psi_N(\{p\})|\Psi_N(\{p\})\rangle.$
Inserting a resolution of identity we obtain
\begin{align}\label{i2}
\langle a_{m_2}(t_2) a_{m_1}^\dagger(t_1)\rangle _N
&=\sum_{\, \{q\}_{N+1}}
\frac{\langle\Psi_N(\{p\})|a_{m_2}(t_2)|\Psi_{N+1}(\{q\})\rangle\langle\Psi_{N+1}(\{q\})| a_{m_1}^\dagger(t_1)|\Psi_N(\{p\})\rangle}{\langle\Psi_N(\{p\})|\Psi_N(\{p\})\rangle
\langle\Psi_{N+1}(\{q\})|\langle\Psi_{N+1}(\{q\}\rangle)}\, ,\nonumber\\
&=\sum_{\, \{q\}_{N+1}}\frac{\ov\F_N(m_2,t_2,\{q\},\{p\})\F_N(m_1,t_1,\{q\},\{p\})}{L^{2N+1}}\, ,
\end{align}
where the sum is over all sets of allowed values for the  momenta $\{q\}$ with $\mbox{dim}\{q\}=N+1$,
and we have introduced the form factors
(note that $\langle\Psi_{N}(\{p\})|a_m(t)|\Psi_{N+1}(\{q\}\rangle=\ov\F_N(m,t,\{q\},\{p\})\, $)
\begin{align}\label{i3}
\F_N(m,t,\{q\},\{p\})&\equiv\langle\Psi_{N+1}(\{q\})|a_m^\dagger(t)|\Psi_N(\{p\}\rangle\, .
\end{align}
Even though the summation appearing in Eq.~(\ref{i2}) seems daunting, as we will
show in the next sections, it can be done exactly in the form of a finite size determinant.
Using this representation in Eq.~(\ref{i1}) the thermodynamic limit can be performed
explicitly obtaining the representations given in Section \ref{Sresults}. Summarizing,
the derivation of the Fredholm determinant representation involves three steps: a) calculation of the form factors
on the finite lattice, b) derivation of a determinant formula for the normalized
mean value of bilocal operators via summation of the form factors, and c) taking the
thermodynamic limit.

\subsection{Form factors on the finite lattice}

The implementation of the program sketched above starts with the calculation
of the form factors. Using $a_m^\dagger(t)=e^{i Ht} a_m^\dagger e^{-i H t}$ we have
\be\label{i4}
\F_N(m,t,\{q\},\{p\})=e^{i t\left( \sum_{a=1}^{N+1}\varepsilon(q_a)- \sum_{b=1}^N\varepsilon(p_b)\right)} \F_N(m,\{q\},\{p\})\, ,
\ee
where we have introduced the static form-factor
$
\F_N(m,\{q\},\{p\})=\langle\Psi_{N+1}(\{q\})|a_m^\dagger|\Psi_N(\{p\}\rangle\, ,
$
which in terms of the wavefunctions can be written as (see Appendix \ref{Aff})
\be\label{i6}
\F_N(m,\{q\},\{p\})=\sqrt{N+1}\sum_{m_1=1}^L\cdots\sum_{m_N=1}^L\ov\chi_{N+1}(m_1,\cdots,m_N,m)|\{q\})
                     \chi_N(m_1,\cdots,m_N|\{p\})\, .
\ee
In Eq.~(\ref{i6})  $\{q\}=q_1,\cdots,q_{N+1}\, , \mbox{ card}\{q\}=N+1$ and
$\{p\}=q_1,\cdots,q_{N}\, , \mbox{ card}\{p\}=N$ are the set of momenta
that parameterize the eigenvectors $|\Psi_{N+1}\rangle$ and $ |\Psi_{N}\rangle$.
They satisfy the BAEs (\ref{BAE})
\begin{align*}
e^{i p_a L}&=e^{-i \pi \st (N-1)}\, ,\ \ \ \ \ a=1,\cdots, N\, ,\\
e^{i q_b L}&=e^{-i \pi \st N}\, , \ \ \ \ \ \ \ \ \ \     b=1,\cdots,N+1\, ,
\end{align*}
and the allowed values are
\begin{subequations}\label{i7}
\begin{align}
p_j&=\frac{2 \pi}{L}\left(-\frac{L}{2}+j\right)+\frac{2\pi \delta}{L}\, , \ \ \delta=\{[-\pi\st(N-1)]\}\, ,\ \ \ j=1,\cdots, L\\
q_l&=\frac{2 \pi}{L}\left(-\frac{L}{2}+l\right)+\frac{2\pi \delta'}{L}\, , \ \ \delta'=\{[-\pi\st N]\}\, ,\ \ \ \ \ \ \ \ \  l=1,\cdots, L\, .
\end{align}
\end{subequations}
Two consequences of (\ref{i7}) which will play an important role in the following
are that $q_b-p_a=\frac{2\pi}{L}(l-\st/2)$ with $l$ integer and $q_b\ne p_a$ except
in the case of spinless fermions ($\st=0$).
Using the explicit expressions of the wavefunctions Eq.~(\ref{i6}) can be rewritten as
\begin{align*}
\F_N(m,\{q\},\{p\})&=\frac{(-i)^N}{N!}\sum_{m_1,\cdots,m_N=1}^L\sum_{Q\in S_{N+1}}\sum_{P\in S_N}(-1)^{[Q]+[P]}e^{-i m q_{Q(N+1)}}
\prod_{a=1}^N\left( e^{-i m_a(q_{Q(a)}-p_{P(a)})-i\pi\st \epsilon(m_a-m)/2}\, ,\right)\\
&=\frac{1}{N!}\sum_{Q\in S_{N+1}}\sum_{P\in S_N}(-1)^{[Q]+[P]}e^{-i m q_{Q(N+1)}}
\prod_{a=1}^N\left((-i)\sum_{m_a=1}^L e^{-i m_a(q_{Q(a)}-p_{P(a)})-i\pi\st \epsilon(m_a-m)/2}\right)\, .
\end{align*}
The summation over $m_i$'s can be done using $e^{-i \pi\st\epsilon(n-m)/2}=\cos(\pi\st/2)-i\sin(\pi\st/2)\epsilon(n-m)$
and the BAEs, $e^{i(q_a-p_b)L}=e^{- i\pi \st},$ with the result
\[
-i\sum_{n=1}^Le^{-i n(q_a-p_b)-i\pi\st\epsilon(n-m)/2}=i \sin\left(\frac{\pi\st}{2}\right)\cot\frac{1}{2}(q_a-p_b)e^{-i m(q_a-p_b)}\, .
\]
Therefore,
\[
\F_N(m,\{q\},\{p\})=\frac{(i \sin(\pi\st/2))^N}{N!}e^{-i m(\sum_{a=1}^{N+1} q_a-\sum_{b=1}^N p_b)}\sum_{P\in S_N} (-1)^{[P]}
\left|
\begin{array}{cccc}
\cot\frac{1}{2}(q_1-p_{P(1)})& \cdots & \cot\frac{1}{2}(q_1-p_{P(N)}) & 1\\
\vdots & \vdots & \vdots &\vdots\\
\cot\frac{1}{2}(q_{N+1}-p_{P(1)})& \cdots & \cot\frac{1}{2}(q_{N+1}-p_{P(N)}) & 1\\
\end{array}
\right|\, .
\]
It is easy to see now that the sum over the  permutations $P$ gives $N!$ identical terms because for each $P$
we can permute the first $N$ columns of the determinant with $P^{-1}$ which multiplies the determinant with
$(-1)^{[P^{-1}]}=(-1)^{[P]}$ canceling the signature of the permutation. Using this observation together with
(\ref{i4}) we obtain the final expression for the form-factor
\be\label{ff}
\F_N(m,t,\{q\},\{p\})=\left(i \sin(\pi\st/2)\right)^N
e^{it\left(\sum_{a=1}^{N+1}\varepsilon(q_a)-\sum_{b=1}^N \varepsilon(p_b)\right)-i m\left(\sum_{a=1}^{N+1}q_a-\sum_{b=1}^N p_b\right)}\,
\mbox{det}_{N+1}\textsf{B} \, ,
\ee
where the $(N+1)\times(N+1)$ matrix $\textsf{B}$ has the elements
\be
\textsf{B}_{ab}=\left\{\begin{array}{ll} \cot\frac{1}{2}(q_a-p_b)\, & a=1,\cdots,N+1, \ \ b=1,\cdots,N\, ,\\
                                          1\, \ \ \ \ \  & a=1,\cdots,N+1, \ \ b=N+1\, .\\
                         \end{array}\right.
\ee
For $\st=1$ the expression for the form factors (\ref{ff}) reduces to the one obtained by
the authors of \cite{CIKT} in their study of the XX0 spin chain.
Subtracting the last row of $\textsf{B}$ from all the other rows, operation which
leaves the determinant unchanged, we obtain
$
\mbox{det}_{N+1}\textsf{B}=\mbox{det}_N(\textsf{A}^{(1)}-z\textsf{A}^{(2)})|_{z=1}\, ,
$
with $\textsf{A}^{(1)}_{ab}=\cot\frac{1}{2}(q_a-p_b)\, ,\  a,b=1,\cdots, N\, ,$ and
$\textsf{A}^{(2)}_{ab}=\cot\frac{1}{2}(q_{N+1}-p_b)\, ,\  a,b=1,\cdots, N\, .$ Another
useful expression
$
\mbox{det}_{N+1}\textsf{B}=\left(1+\frac{d}{dz}\right)\mbox{det}_{N}\textsf{A}|_{z=0}\, ,
$
can be derived using the
fact that $\mbox{det}_N \textsf{A}=\mbox{det}_N(\textsf{A}^{(1)}-z\textsf{A}^{(2)})$
is  a linear function in $z$ (the rank of $\textsf{A}^{(2)}$ is 1)
and the identity   $\left(1+\frac{d}{dz}\right)f(z)|_{z=0}=f(1)$ valid for a linear function  $f(z)=a+bz.$

\subsection{Normalized mean value of $\langle a_{m_2}(t_2) a_{m_1}^\dagger(t_1)\rangle_N$}

Having derived a compact expression for the form factors now we can compute normalized mean values
of bilocal operators. We will consider first  $\langle a_{m_2}(t_2) a_{m_1}^\dagger(t_1)\rangle_N$.
Starting from Eq.~(\ref{i2}) and using (\ref{ff}) we find
\begin{align}\label{s1}
\langle a_{m_2}(t_2) a_{m_1}^\dagger(t_1)\rangle_N
&=\frac{(\sin(\pi\st/2))^{2N}}{L^{2N+1}}\sum_{q_1<\cdots<q_{N+1}}
e^{-it\left(\sum_{a=1}^{N+1}\varepsilon(q_a)-\sum_{b=1}^N \varepsilon(p_b)\right)+i m\left(\sum_{a=1}^{N+1}q_a-\sum_{b=1}^N p_b\right)}
\left(\mbox{det}_{N+1}\textsf{B}\right)^2\, .
\end{align}
In the following it will be useful to introduce the function
\be
e_-(m,t,p)\equiv e_-(p)=e^{-im p/2-iJt\cos p}\, ,
\ee
in terms of which the exponential appearing in the r.h.s of (\ref{s1})
denoted by $f(m,t,\{q\},\{p\})\equiv f(\{q\},\{p\})$ is given by
\[
f(\{q\},\{p\})=e^{-2 i h t}\prod_{a=1}^{N+1}\left(e_-(m,t,q_a)\right)^{-2}\prod_{b=1}^{N}\left(e_-(m,t,p_b)\right)^2\, .
\]
An important observation that we make is that $\left(\mbox{det}_{N+1}\textsf{B}\right)^2$ is a symmetric
function of $q$'s which vanishes when two $q$'s coincide. This means that we can write
\be\label{sum}
\sum_{q_1<\cdots<q_{N+1}}=\frac{1}{(N+1)!}\sum_{q_1}\cdots\sum_{q_{N+1}}\, ,
\ee
where the summation over $q$ for an arbitrary function $\phi$ is given by
$ \sum_{q_a} \phi(q_a)=\sum_{l=1}^L\phi([q_a]_l)\, ,$ with $
[q_a]_l=\frac{2\pi}{L}\left(-\frac{L}{2}+l\right)+\frac{2\pi \delta'}{L}\, ,\ \ l=1,\cdots,L\, .$
Using the definition of the determinant $\mbox{det}_N\textsf{B}=\sum_{Q\in S_N}(-1)^{[Q]}
\prod_{a=1}^{N}\cot\frac{1}{2}(q_{Q(a)}-p_a)$ and (\ref{sum}) we find
\begin{align}\label{i8}
\langle &a_{m_2}(t_2)  a_{m_1}^\dagger(t_1)\rangle_N= \nonumber\\
&=\frac{(\sin(\pi\st/2))^{2N}}{(N+1)!\, L^{2N+1}}
\sum_{q_1,\cdots,q_{N+1}}f(\{q\},\{p\})\sum_{P,Q\in S_{N+1}}(-1)^{[P]+[Q]}\prod_{a=1}^N
\cot\frac{1}{2}(q_{P(a)}-p_a)\cot\frac{1}{2}(q_{Q(a)}-p_a)\nonumber\\
&=\frac{(\sin(\pi\st/2))^{2N}}{(N+1)!\, L^{2N+1}}
\sum_{q_1,\cdots,q_{N+1}}f(\{q\},\{p\})\sum_{Q,R\in S_{N+1}}(-1)^{[R]+[Q]+[Q]}\prod_{a=1}^N
\cot\frac{1}{2}(q_{RQ(a)}-p_a)\cot\frac{1}{2}(q_{Q(a)}-p_a)    \nonumber\\
&=\frac{(\sin(\pi\st/2))^{2N}}{(N+1)!\, L^{2N+1}}
\sum_{q_1,\cdots,q_{N+1}}f(\{q\},\{p\})\sum_{Q\in S_{N+1}}
\left|
\begin{array}{cccc}
\cot\frac{1}{2}(q_{Q(1)}-p_1)& \cdots & \cot\frac{1}{2}(q_{Q(1)}-p_{N}) & 1\\
\vdots & \vdots & \vdots &\vdots\\
\cot\frac{1}{2}(q_{Q(N+1)}-p_{1})& \cdots & \cot\frac{1}{2}(q_{Q(N+1)}-p_{N}) & 1\\
\end{array}
\right|
 \prod_{a=1}^N \cot\frac{1}{2}(q_{Q(a)}-p_{a})\, ,
\end{align}
where in the second line we have used the fact that for a given permutation $Q$
every permutation $P$ can be written as $P=RQ$. Multiplying the $i$-th row (column)
($i=1,\cdots,N$) of the the determinant appearing in
(\ref{i8})  with $e_-(p_i)\cot\frac{1}{2}(q_{Q(i)}-p_i)/e_-^2(q_{Q(i)})$
($e_-(p_i)\sin^2(\pi\st/2)/L^2$) and the $N+1$-th row (column) with
$1/e_-^2(q_{Q_{N+1}})$ ($1/L$) then $q_{Q_i}$ appears only in
the $i$-th row which means that we can sum over $q$'s inside the determinant. Also the
sum over permutation gives $(N+1)!$ identical terms. Introducing the functions
\begin{subequations}
\begin{align}
\U_{ab}&=\frac{\sin^2(\pi\st/2)}{L^2}\sum_qe^{i m q+2 i J t \cos q}\cot\frac{1}{2}(q-p_a)\cot\frac{1}{2}(q-p_b)\, ,\\
g(m,t)&=\frac{1}{L}\sum_qe^{i m q+2 i J t \cos q}\, ,\\
e(m,t,p)\equiv e(m)&=\frac{1}{L}\sum_qe^{i m q+2 i J t \cos q}\cot\frac{1}{2}(q-p)\, ,\\
e_+(m,t,p)\equiv e_+(m)&=e_-(m,t,p)e(m,t,p)\, ,
\end{align}
\end{subequations}
we find
\begin{align}\label{i8b}
\langle a_{m_2}(t_2) & a_{m_1}^\dagger(t_1)\rangle_N = e^{-2 i h t}
\left|
\begin{array}{llll}
\U_{11}e_-(p_1)e_-(p_1)& \cdots & \U_{1N}e_-(p_1)e_-(p_N) & e_+(p_1)\\
\vdots & \vdots & \vdots & \vdots\\
\U_{1N}e_-(p_N)e_-(p_1)& \cdots & \U_{NN}e_-(p_N)e_-(p_N) & e_+(p_N)\\
e_+(p_1)\sin^2(\pi\st/2)/L& \cdots & e_+(p_N)\sin^2(\pi\st/2)/L & g(m,t)
\end{array}
\right|\, .
\end{align}
Expanding on the last column we obtain
\[
\langle a_{m_2}(t_2)  a_{m_1}^\dagger(t_1)\rangle_N = e^{-2 i h t}\left[g(m,t)+\frac{d}{dz}\right]
\mbox{det}_N\left.\left(\U_{ab}e_-(p_a)e_-(p_b)-z \R_{ab}^{(+)}\right)\right|_{z=0}\, ,
\]
where we have introduced
\be\label{rp}
\R_{ab}^{(+)}=\sin^2(\pi\st/2)e_+(p_a)e_+(p_b)/L\, .
\ee
Making use of the identity $\cot\frac{1}{2}(q-p_a)\cot\frac{1}{2}(q-p_b)=\cot\frac{1}{2}(p_a-p_b)
\left[\cot\frac{1}{2}(q-p_a)-\cot\frac{1}{2}(q-p_b)\right]-1$
and introducing
\be\label{d}
d(m,t,p,\st)\equiv d(p)=\frac{\sin^2(\pi\st/2)}{L^2}\sum_q \frac{e^{i m q+2 i J t \cos q}}{\sin^2\frac{1}{2}(q-p)}\, ,\\
\ee
we obtain the final result for the  normalized value of $\langle a_{m_2}(t_2) a^\dagger_{m_1}(t_1)\rangle_N$
on the finite lattice
\be\label{gplusf}
\langle a_{m_2}(t_2)  a_{m_1}^\dagger(t_1)\rangle_N = e^{-2 i h t}\left[g(m,t)+\frac{d}{dz}\right]
\mbox{det}_N\left.\left(\textsf{S}_{ab}-z \R_{ab}^{(+)}\right)\right|_{z=0}\, ,
\ee
with the elements of the matrix $\textsf{S}$  given by
\be\label{defS}
\textsf{S}_{ab}=\delta_{ab}d(p_a)e_-^2(p_a)+(1-\delta_{ab})\sin^2(\pi\st/2)
\frac{e_+(p_a)e_-(p_b)-e_-(p_a)e_+(p_b)}{L\tan\frac{1}{2}(p_a-p_b)}
-\frac{\sin^2(\pi\st/2)}{L}g(m,t)e_-(p_a)e_-(p_b)\, .
\ee

\subsection{Normalized mean value of $\langle a_{m_2}^\dagger(t_2) a_{m_1}(t_1)\rangle_N$}

The derivation of a determinant representation for $\langle a_{m_2}^\dagger(t_2) a_{m_1}(t_1)\rangle_N$
is very similar with the one presented in the previous section. The main difference is that compared with
Eq.~(\ref{i2}) the resolution of identity contains now eigenstates with $N-1$ particles. Therefore
\begin{align}\label{i9}
\langle a_{m_2}^\dagger(t_2) a_{m_1}(t_1)\rangle_N
&=\sum_{q_1<\cdots<q_{N-1}}\frac{\F_{N-1}(m_2,t_2,\{p\},\{q\})\ov\F_{N-1}(m_1,t_1,\{p\},\{q\})}{L^{2N-1}}\, ,\nonumber\\
&=\frac{(\sin(\pi\st/2))^{2N-2}}{L^{2N-1}}\sum_{q_1<\cdots<q_{N-1}}
e^{it\left(\sum_{a=1}^{N}\varepsilon(p_a)-\sum_{b=1}^{N-1} \varepsilon(q_b)\right)-i m\left(\sum_{a=1}^{N}p_a-\sum_{b=1}^{N-1} q_b\right)}
\left(\mbox{det}_{N}\textsf{B}\right)^2\, .
\end{align}
In the last line we have used that $\F_{N-1}(m,t,\{p\},\{q\})\equiv \langle \Psi_N(\{p\}|a_m^\dagger(t)|\Psi_{N-1}(\{q\})\rangle$
(note the interchange of  $\{p\}$ and $\{q\}$) is given by
\begin{align*}
\F_{N-1}(m,t,\{p\},\{q\})
&=(i \sin(\pi\st/2))^{N-1}
e^{it\left(\sum_{a=1}^{N}\varepsilon(p_a)-\sum_{b=1}^{N-1} \varepsilon(q_b)\right)-i m\left(\sum_{a=1}^{N}p_a-\sum_{b=1}^{N-1} q_b\right)}
\mbox{det}_{N} \textsf{B}\, ,
\end{align*}
with $\mbox{card}\{p\}=N, $  $\mbox{card}\{q\}=N-1$ and the $N\times N$ matrix $\textsf{B}$ has the elements
\be
\textsf{B}_{ab}=\left\{\begin{array}{ll} \cot\frac{1}{2}(p_a-q_b)\, & a=1,\cdots,N, \ \ b=1,\cdots,N-1\, ,\\
                                          1\, \ \ \ \ \  & a=1,\cdots,N, \ \ b=N\, .\\
                         \end{array}\right.
\ee
%
Similar to the case treated in the previous section $\left(\mbox{det}_{N}\textsf{B}\right)^2$ is a symmetric function of  $q$'s and vanishes
when two of them coincide. Replacing the sum appearing in (\ref{i9}) with
\[
 \sum_{q_1<\cdots<q_{N-1}}=\frac{1}{(N-1)!}\sum_{q_1}\cdots\sum_{q_{N-1}}, ,
\]
and using
$\mbox{det}_N\textsf{B}=\sum_{Q\in S_N}(-1)^{[Q]}\prod_{a=1}^{N-1}\cot\frac{1}{2}(p_{Q(a)}-q_a)$
we find
\begin{align}\label{i10}
\langle &a_{m_2}^\dagger(t_2)  a_{m_1}(t_1)\rangle_N = \nonumber\\
&\frac{(\sin(\pi\st/2))^{2(N-1)}}{(N-1)!\, L^{2N-1}}
\sum_{q_1,\cdots,q_{N-1}}f(\{p\},\{q\})\sum_{P,Q\in S_{N}}(-1)^{[P]+[Q]}\prod_{a=1}^{N-1}
\cot\frac{1}{2}(p_{P(a)}-q_a)\cot \frac{1}{2}(p_{Q(a)}-q_a)\nonumber\\
&\frac{(\sin(\pi\st/2))^{2(N-1)}}{(N-1)!\, L^{2N-1}}
\sum_{q_1,\cdots,q_{N-1}}f(\{p\},\{q\})\sum_{Q,R\in S_{N}}(-1)^{[R]+[Q]+[Q]}\prod_{a=1}^{N-1}
\cot\frac{1}{2}(p_{RQ(a)}-q_a)\cot\frac{1}{2}(p_{Q(a)}-q_a)   \nonumber\\
&\frac{(\sin(\pi\st/2))^{2(N-1)}}{(N-1)!\, L^{2N-1}}
\sum_{q_1,\cdots,q_{N-1}}f(\{p\},\{q\})\sum_{Q\in S_{N}}
\left|
\begin{array}{cccc}
\cot\frac{1}{2}(p_{Q(1)}-q_1)& \cdots & \cot \frac{1}{2}(p_{Q(1)}-q_{N-1}) & 1\\
\vdots & \vdots & \vdots &\vdots\\
\cot\frac{1}{2}(p_{Q(N)}-q_{1})& \cdots & \cot\frac{1}{2}(p_{Q(N)}-q_{N-1}) & 1\\
\end{array}
\right|
 \prod_{a=1}^{N-1} \cot\frac{1}{2}(p_{Q(a)}-q_{a})\, ,
\end{align}
where
$
f(\{p\},\{q\})=e^{2i t h}\prod_{a=1}^N(e_-(m,t,p_a))^2\prod_{b=1}^{N-1}(e_-(m,t,q_b))^{-2}\, .
$
Multiplying the $i$-th row  ($i=1,\cdots,N$) of the the determinant appearing in
(\ref{i10})  with $e_-(p_{Q(i)})$ the $j$-th colum ($j=1,\cdots, N-1$) with
$\sin^2(\pi\st/2)\cot\frac{1}{2}(p_{Q(j)}-q_j)e_-(p_{Q(j)})/(L^2 e_-^2(q_j))$ and the
$N$-th column with $e_-(p_{Q(N)})/L$ and summing over $q$'s inside the determinant
(this is allowed because $q_i$ appears only in the $i$-th column) we obtain
\begin{align*}
\langle a_{m_2}^\dagger(t_2)&  a_{m_1}(t_1)\rangle_N =\nonumber\\
&\frac{ e^{ 2i t h}}{(N-1)!}\sum_{Q\in S_N}
\left|
\begin{array}{llll}
\U_{Q(1)Q(1)}e_-(p_{Q(1)})e_-(p_{Q(1)})& \cdots & \U_{Q(1)Q(N-1)}e_-(p_{Q(1)})e_-(p_{Q(N-1)}) & \R^{(-)}_{Q(1)Q(N)}\\
\ \ \ \ \ \ \ \ \ \ \ \vdots & \ \, \vdots & \ \ \ \ \ \ \ \ \ \ \vdots &\ \ \ \ \   \vdots\\
\U_{Q(N)Q(1)}e_-(p_{Q(N)})e_-(p_{Q(1)})& \cdots & \U_{Q(N)Q(N-1)}e_-(p_{Q(N)})e_-(p_{Q(N-1)}) & \R^{(-)}_{Q(N)Q(N)}\\
\end{array}
\right|\, ,
\end{align*}
where we have introduced
\be\label{rm}
\R_{ab}^{(-)}=\frac{e_-(p_a)e_-(p_b)}{L}\, .
\ee
Permuting the rows and columns with $Q^{-1}$, operation which leaves the determinant unchanged, the sum
over permutations gives $(N-1)!$ identical terms. The final result is
\begin{align}\label{gm}
\langle a_{m_2}^\dagger(t_2)  a_{m_1}(t_1)\rangle_N &= e^{ 2it h} \frac{d}{dz}\left.\mbox{det}_N(\textsf{S}+z\R^{(-)})\right|_{z=0}
=e^{ 2it h}[\mbox{det}_N(\textsf{S}+\R^{(-)})-\mbox{det}_N\textsf{S}]\, .
\end{align}
with $\textsf{S}$ defined by (\ref{defS}).

\subsection{Thermodynamic limit of the correlators}\label{Sther}

The final step in deriving the results presented in Section \ref{Sresults} involves
taking the thermodynamic limit in the finite lattice results obtained in the
previous sections. Before we do that we remind the reader that the Fredholm determinant (see Chap. XI of \cite{WW})
of  an integral operator $\hat K$ which acts on an arbitrary function as
$
(\hat K f)(p)=\int_a^b K(p,q)f(q) dq\,
$
is an entire function of $\gamma$ defined as
\begin{align}
\det (1-\gamma\hat K)&=\lim_{n\rightarrow\infty}\left\{
1-\gamma\sum_{j_1=1}^n \delta K(p_{j_1},p_{j_1})+\frac{\gamma^2}{2!}
\sum_{j_1,j_2=1}^n\delta^2\left|\begin{array}{ll}
K(p_{j_1},p_{j_1}) & K(p_{j_1},p_{j_2})\\
K(p_{j_2},p_{j_1}) & K(p_{j_2},p_{j_2})
\end{array}
\right|+\cdots
\right\}\, ,\nonumber\\
&=
1-\gamma\int_a^b  K(p_1,p_1) dp_1+\frac{\gamma^2}{2!}
\int_a^b\int_a^b \left|\begin{array}{ll}
K(p_1,p_1) & K(p_1,p_2)\\
K(p_2,p_1) & K(p_2,p_2))
\end{array}
\right| dp_1dp_2+\cdots
\, .
\end{align}
Let us start with $\gp^{(T)}_+(m,t).$ The thermodynamic limit of this correlator is
\be
\gp_+^{(T)}(m,t)=\lim_{L\rightarrow\infty}\frac{\sum_{N=0}^L\sum_{p_1<\cdots<p_N}e^{-\frac{E_N(\{p\})}{T}}
\langle a_{m_2}(t_2)a_{m_1}^\dagger(t_1)\rangle_N}{\textsf{tr }[e^{-H/T}]}\, .
\ee
with $\langle a_{m_2}(t_2)a_{m_1}^\dagger(t_1)\rangle_N$ given by (\ref{gplusf}).
The denominator is
\begin{align}
\textsf{tr }[e^{-H/T}]&=\sum_{N=0}^L\sum_{p_1<\cdots<p_N}e^{-\frac{E_N(\{p\})}{T}}
=\prod_{p}\left(1+e^{-\varepsilon(p)/T}\right)\, , \nonumber\\
&\simeq \exp\left\{\frac{L}{2\pi}\intp \ln(1+e^{-\varepsilon(p)/T}) dp\right\}\, .
\end{align}
with $\varepsilon(p)=-2J\cos p+2h$ and $L\rightarrow\infty.$ This expression is divergent.
A Fredholm determinant is well defined if the trace of the operator  is finite $\inti K(p,p) dp<\infty $.
Fortunately, as we will see below this divergent expression can be extracted from the numerator
making the expression for the correlator well defined. We can write that
\[
\textsf{tr }[e^{-H/T}]=\det(1+\hat{\textsf{Z}})\, ,\ \  \textsf{Z}(p,p')=e^{-\varepsilon(p)/T}\delta_L(p-p')\, ,
\]
with $\delta_L(p-p')=\frac{\sin(L(p-p'))}{2 \pi(p-p')}$ is a regularization of the delta function.
The numerator can be written as
\begin{align}
\lim_{L\rightarrow\infty}&\sum_{N=0}^L\sum_{p_1<\cdots<p_N}e^{-\frac{E_N(\{p\})}{T}}
\langle a_{m_2}(t_2)a_{m_1}^\dagger(t_1)\rangle_N=
\lim_{L\rightarrow\infty}\sum_{N=0}^L\frac{1}{N!}\sum_{p_1}\cdots\sum_{p_N}e^{-\frac{\sum_{a=1}^N\varepsilon(p_a))}{T}}
\langle a_{m_2}(t_2)a_{m_1}^\dagger(t_1)\rangle_N\, ,\nonumber\\
&=\lim_{L\rightarrow\infty}\sum_{N=0}^L\frac{1}{N!}\sum_{p_1}\cdots\sum_{p_N}
e^{-2 i h t}\left[g(m,t)+\frac{\6}{\6 z}\right]\left.\mbox{det}_N(\tilde{\textsf{S}}-z \tilde{\R}^{(+)})\right|_{z=0}\, ,\nonumber\\
&=\lim_{L\rightarrow\infty}\sum_{N=0}^L\frac{1}{N!}\sum_{p_1}\cdots\sum_{p_N}
e^{-2 i h t}\left[(g(m,t)-1)\mbox{det}_N\tilde{\textsf{S}}+\mbox{det}_N(\tilde{\textsf{S}}- \tilde{\R}^{(+)})\right]\, ,
\end{align}
with $\tilde{\textsf{S}}$ and $\tilde{\R}^{(+)}$ obtained from (\ref{defS}) and (\ref{rp}) via the
transformations
$\tilde{\textsf{S}}_{ab}=e^{-\varepsilon(p_a)/2T}\, \textsf{S}_{ab}\,  e^{-\varepsilon(p_b)/2T}\, ,$
$\tilde{\R}^{(+)}_{ab}=e^{-\varepsilon(p_a)/2T}\, \R^{(+)}_{ab}\,  e^{-\varepsilon(p_b)/2T}\, .$
Performing the thermodynamic limit and remembering that in this limit $e(p)\rightarrow E(p)$, $g\rightarrow G\, ,$
$d(p)\rightarrow D(p)$ (see (\ref{TLg}), (\ref{TLe}), (\ref{TLd})) we obtain
\begin{align}\label{i30}
\gp_+^{(T)}(m,t)&=\frac{e^{-2 i h t}}{\mbox{det}(1+\hat{\textsf{Z}})}
\left[G(m,t)+\frac{\6}{\6 z}\right]\left.\mbox{det}
(1+\hat{\textsf{Z}}+\hat{\tilde{\textsf{V}}}-z \hat{\tilde{\R}}^{(+)})\right|_{z=0}\, ,\nonumber\\
&=\frac{e^{-2 i h t}}{\mbox{det}(1+\hat{\textsf{Z}})}
\left[(G(m,t)-1)\mbox{det}(1+\hat{\textsf{Z}}+\hat{\tilde{\textsf{V}}})+
\mbox{det}(1+\hat{\textsf{Z}}+\hat{\tilde{\textsf{V}}}-\hat{\tilde{\R}}^{(+)}) \right]\, ,
\end{align}
with kernels
\begin{align}
\tilde V(p,p')&=\sin^2\left(\frac{\pi\st}{2}\right)e^{-\varepsilon(p)/2T} e^{-\varepsilon(p')/2T}\left[\frac{E_+(p)E_-(p')-E_-(p)E_+(p')}{\tan\left[(p-p')/2\right]}-G(m,t)E_-(p)E_-(p')\right]\, ,\nonumber\\
\tilde R^{(+)}(p,p')&=\sin^2\left(\frac{\pi\st}{2}\right)e^{-\varepsilon(p)/2T} e^{-\varepsilon(p')/2T}E_+(p)E_+(p')\, .
\end{align}
It should be noted that the second term on the r.h.s of (\ref{TLd})) is obtained from the first term appearing
in the square bracket of $\tilde V(p,p')$ in the limit $p\rightarrow p'$.
Extracting $\mbox{det}(1+\hat{\textsf{Z}})$ from the numerator of (\ref{i30}) which cancels the similar term in the
denominator we obtain the Fredholm determinant representation of $\gp_+^{(T)}(m,t)$ (\ref{gplus}). The thermodynamic
limit of $\gp_-^{(T)}(m,t)$ is performed along similar lines with the result (\ref{gminus}).

\section{Conclusions}\label{Sconc}

We have derived Fredholm determinant representations for the time-, space-, and temperature-dependent
Green's function of a system of 1D hard-core anyons which can be understood as the fractional
statistics generalization  of hard-core bosons (XX0 spin chain). In the static case we have also computed
the large distance asymptotics and the momentum distribution function at zero and finite temperature.
The results obtained in this paper can be used to rigorously
investigate the time-dependent correlation functions as in
the case of the XX0 spin chain \cite{IIKS}. The first step would be the derivation
of the classical integrable  system of differential equations characterizing the
correlators. Based on similar results for Lieb-Liniger anyons \cite{SC,L3,L4} we expect to
obtain the same Ablowitz-Ladik system which characterizes the  XX0 spin chain correlators
\cite{IIKS} but  with different boundary conditions.  We should also point out that even
though it is not as straightforward as in the static case, dynamic correlators can also
be numerically investigated using the methods of \cite{Bor}. This is left for further research.

\section{Acknowledgments}

Financial support from  the PNII-RU-TE-2012-3-0196 grant  of the Romanian National Authority for
Scientific Research is gratefully acknowledged.

\appendix

\section{Finite size corrections}\label{Afsc}

In this appendix we  derive the finite size corrections of the low-lying
excitations. In order to make the discussion below
as simple and transparent as possible we consider the number of particles
in the ground state to be $N$ with $N$ odd such that $e^{i \pi (1-\st)(N-1)}=1$. With this
condition the BAEs for systems with $N$ and $N+1$ particles are $e^{i p_a L}=e^{-i \pi \st(N-1)}=(-1)^{N-1}\, ,\ a=1,\cdots,N$
and $e^{i q_a L}=e^{-i \pi \st N}=(-1)^{N} e^{i\pi(1-\st)}\, ,\ a=1,\cdots,N+1$.
The momenta of the particles in the groundstates are
\begin{subequations}\label{qs}
\begin{align}
p_j&=\frac{2\pi}{L}\left[-\frac{N+1}{2}+j\right]\, ,\ \  j=1,\cdots,N\, ,\\
q_j&=\frac{2\pi}{L}\left[-\frac{N+2}{2}+\frac{1-\st}{2}+j\right]\, ,\ \  j=1,\cdots,N+1\, .
\end{align}
\end{subequations}
We make a simple observation which will play an important role in the following
discussion. We notice that BAEs of  the system with $N$ particles are the same as
the BAEs of the XX0 spin chain \cite{CIKT} with the same number of quasiparticles.
Therefore, if the momenta of the quasiparticles in the groundstates of the
XXO spin chain with $N$ and $N+1$ quasiparticles   are denoted by $p_j^0$ and $q_j^0$
we have $q_j=q_j^0\, ,\ p_j=p_j^0+\frac{2\pi}{L}\frac{1-\st}{2}$.

In the thermodynamic limit at zero temperature we have $p_{j+1}-p_j=q_{j+1}-q_{j}=\frac{2\pi}{L}$ and
the momenta fill densely the interval $[-k_F,k_F]$ which means that we can
replace sums with integrals using the rule $\sum\rightarrow \frac{L}{2\pi}\int_{-k_F}^{k_F}$.
Due to the fact that the groundstate with $N$ particles is identical with the XX0 spin
chain case we have $E(N,L)-L e_0=-\frac{\pi v_F}{6L}c+O(1/L^2)$ with $c=1$ the central charge
and $v_F=\varepsilon'(k_F)=2J \sin k_F$ the Fermi velocity. Now we are ready to compute
the finite size corrections.

{\it Addition of one particle into the system.} The momenta characterizing the groundstates
of systems with $N$ and $N+1$ particles are given by (\ref{qs}). The momentum of this excitation is easily
obtained  $\Delta P (\Delta N=1)=2 k_F(\frac{1-\st}{2})+\frac{2\pi}{L}(\frac{1-\st}{2})$
due to $\sum p_j^0=\sum q_j^0=0$ and $k_F=\pi D$ with $D=N/L$. Introducing the
notation $\omega=\frac{2\pi}{L}\frac{1-\st}{2}$ the energy of the excitation is
\begin{align*}
\Delta E(\Delta N=1)&=\sum_{j=1}^{N+1}\varepsilon(q_j)-\sum_{j=1}^{N}\varepsilon(p_j)\, , \\
&=\sum_{j=1}^{N+1}\left[\varepsilon(q_j^0)+\varepsilon'(q_j^0)\omega+\varepsilon''(q_j^0)\frac{\omega^2}{2}
\right]-\sum_{j=1}^{N}\varepsilon(p_j^0)\, , \\
&=\Delta E^0(\Delta N=1)+\frac{L}{2\pi}\int_{-k_F}^{k_F}  \varepsilon'(q)\omega+\varepsilon''(q)\frac{\omega^2}{2} \, dq\, ,\\
&=\frac{2\pi v_F}{L}\left[ \left(\frac{1}{2}\right)^2+\left(\frac{1-\st}{2}\right)^2\right]
\end{align*}
where we have used the finite size correction formula for the XXO spin chain (Chap. II of \cite{KBI}
with $\mathcal{Z}=1$) $\Delta E^0(\Delta N=1)=\frac{\pi v_F}{2L}$  and the fact that $\varepsilon'(q)$
is an odd function which gives zero after the integration over a symmetric interval.

{\it Particle-hole excitation.} We consider first a particle-hole excitation at the right Fermi boundary $k_F$.
The momenta are still given by (\ref{qs}) with the exception of $\tilde {q}_{N+1}$ which is
$\tilde {q}_{N+1}=q_{N+1}+\frac{2\pi }{L}N^+$ with $N^+$ an arbitrary integer. The momentum of the excitation  is simply
$\Delta P(N^+)=\frac{2\pi N^+}{L}$ and the energy $\Delta E(N^+)=\varepsilon(\tilde {q}_{N+1})-\varepsilon({q}_{N+1})=
\frac{2 \pi v_F}{L}N^+$ using $\varepsilon(\tilde {q}_{N+1})=\varepsilon({q}_{N+1})+\varepsilon'({q}_{N+1})\frac{2 \pi }{L}N^+$.
Similar formulas, with $N^+$ replaced by $N^-$, can be derived in the case of a  particle-hole excitation at
the left Fermi boundary $-k_F$.

{\it Backscattering of $d$ particles.} In this case $d$ particles jump from $-k_F$ to $k_F$. The momenta of the particles
in the excited state are $\tilde q_j=\frac{2}{\pi L}\left[-\frac{N+2}{2}+\frac{1-\st}{2}+d+j\right]=q_j^0+\omega+\omega'$
with $\omega'=\frac{2\pi}{L} d$. The momentum of the excitation is $\Delta P(d) =2k_F d +\frac{2\pi}{L} d$ and
the energy
\begin{align*}
\Delta E(d)&=\sum_{j=1}^{N+1}\varepsilon(\tilde q_j)-\sum_{j=1}^{N+1}\varepsilon(q_j)\, , \\
&=\sum_{j=1}^{N+1}\left[\varepsilon(q_j^0)+\varepsilon'(q_j^0)(\omega+\omega')+\varepsilon''(q_j^0)\frac{(\omega+\omega')^2}{2}
\right]-\sum_{j=1}^{N+1}\left[\varepsilon(q_j^0)+\varepsilon'(q_j^0)\omega+\varepsilon''(q_j^0)\frac{\omega^2}{2}\right]\, , \\
&=\frac{2\pi v_F}{L}\left[ \left(d+\frac{1-\st}{2}\right)^2-\left(\frac{1-\st}{2}\right)^2\right]\, .
\end{align*}
Collecting all the results we obtain Eq.~(\ref{corrections}).

\section{Wavefunctions expression for the form factor}\label{Aff}

In this Appendix we derive Eq.~(\ref{i6}) which is the starting
point in our calculation of the from factors. It is instructive to consider
first the simple case of $N=2$  which contains all the relevant features associated
with anyonic statistics. The generalization to any value of $N$ is straightforward.
Starting with the eigenvectors for two and three particles
\begin{align*}
|\Psi_2(\{p\})\rangle&=\frac{1}{\sqrt{2!}}\sum_{m_1=1}^L\sum_{m_2=1}^L\chi_2(m_1,m_2|(\{p\})a_{m_2}^\dagger a_{m_1}^\dagger|0\rangle\, ,\\
\langle \Psi_3(\{q\})|&=\frac{1}{\sqrt{3!}}\sum_{n_1=1}^L\sum_{n_2=1}^L\sum_{n_3=1}^L\ov\chi_3(n_1,n_2,n_3|\{q\})\langle 0|a_{n_1} a_{n_2} a_{n_3}\, ,
\end{align*}
we obtain for the static form factor the following expression
\be\label{a1}
\F_2(m,\{q\},\{p\})=\frac{1}{\sqrt{2!3!}}\sum_{n_1,n_2,n_3=1}^L\sum_{m_1,m_2=1}^L\ov\chi_3(n_1,n_2,n_3|\{q\})\chi_2(m_1,m_2|(\{p\})
\langle 0|a_{n_1} a_{n_2} a_{n_3}a_m^\dagger a_{m_2}^\dagger a_{m_1}^\dagger|0\rangle\, .
\ee
Moving successively the annihilation operators to the right with the help of the commutation relations (\ref{comm})
and using $a_j|0\rangle=0$ we find
\begin{align*}
\langle 0|a_{n_1} a_{n_2} a_{n_3}&a_m^\dagger a_{m_2}^\dagger a_{m_1}^\dagger|0\rangle=
\delta_{n_3, m}\delta_{n_2,m_2}\delta_{n_1,m_1}-e^{-i \pi \st\epsilon(n_2-m_2)}\delta_{n_3,m}\delta_{n_1,m_2}\delta_{n_2,m_1}
-e^{-i \pi\st\epsilon(n_3-m)}\delta_{n_3,m_2}\delta_{n_2,m}\delta_{n_1,m_1}\\
&+e^{-i\pi\st[\epsilon(n_3-m)+\epsilon(n_2-m)]}\delta_{n_3,m_2}\delta_{n_2,m_1}\delta_{n_1,m}
 +e^{-i\pi\st[\epsilon(n_3-m)+\epsilon(n_3-m_2)]}\delta_{n_3,m_1}\delta_{n_1,m_2}\delta_{n_2,m}\\
&-e^{-i \pi \kappa[\epsilon(n_3-m)+\epsilon(n_3-m_2)+\epsilon(n_2-m)]}\delta_{n_3,m_1}\delta_{n_2,m_2}\delta_{n_1,m}\, .
\end{align*}
Plugging this relation in (\ref{a1}) and using the anyonic symmetry of the wavefunctions (\ref{exchange})
we obtain
\[
\F_2(m,\{q\},\{p\})=\sqrt{3}\sum_{m_1=1}^L\sum_{m_2=1}^L\sum_{m_3=1}^L\ov\chi_3(m_1,m_2,m|\{q\})\chi_2(m_1,m_2|(\{p\})\, .
\]
The generalization of this result in the case of $N$ particles is given  by Eq.~\ref{i6}.

\section{Thermodynamic limit of singular function}

Here we calculate the thermodynamic limit of the functions introduced in Section \ref{Sdet}. On the
finite lattice the  allowed values for $q$'s are given by
$q_l=\frac{2 \pi}{L}\left(-\frac{L}{2}+l\right)+\frac{2\pi \delta'}{L}\, $ with $l=1,\cdots,L\, .$
In the thermodynamic limit  $q_{l+1}-q_l=2 \pi/L$ and the $q$'s fill densely the interval $[-\pi,\pi]$
which means that the sums appearing in the definition of the various functions defined in
Section \ref{Sdet} can be replaced by integrals using the following rule
\be\label{a2}
\frac{1}{L}\sum_q\rightarrow \frac{1}{2\pi}\int_{-\pi}^\pi d\, q\, .
\ee
In the case of functions involving differences of the type $q-p$ we need to distinguish between
the $\langle a_{m_2}(t_2) a_{m_1}^\dagger(t_1)\rangle_N$ and $\langle a^\dagger_{m_2}(t_2) a_{m_1}(t_1)\rangle_N$
cases which  will be denoted by $\langle a a^\dagger\rangle$ and $\langle a^\dagger a\rangle$.
This is because the BAEs  satisfied by the $q$'s are $e^{i q L}=e^{-i \pi\st N}$ ($\langle a a^\dagger\rangle$ case)
and $e^{i q L}=e^{-i \pi\st (N-2)}$ ($\langle a^\dagger a\rangle$ case)  with $e^{i p L}=e^{-i \pi\st (N-1)}$
in both cases. Therefore,
\be
\sum_q f(q-p)=\left\{\begin{array}{l}
\sum_{j=1}^L f[\frac{2\pi}{L}(j-k-\st/2)]\, ,\  k\in\{1,\cdots,L\} \mbox { for } \langle a a^\dagger\rangle\, ,\\
\sum_{j=1}^L f[\frac{2\pi}{L}(j-k+\st/2)]\, ,\  k\in\{1,\cdots,L\} \mbox { for } \langle a^\dagger a\rangle\, .
\end{array}
\right.
\ee
The simplest situation is encountered in the case of the $g(m,t)$ function. Using (\ref{a2}) we
have
\be\label{TLg}
g(m,t)=\frac{1}{L}\sum_{q}e^{ i m q+2i Jt \cos q}
\longrightarrow G(m,t)=\frac{1}{2\pi}\intp dq e^{ i m q+2i Jt \cos q}\, .
\ee
The first nontrivial case is encountered in the case of
$
e(m,t,p)=\frac{1}{L}\sum_q\frac{e^{ i m q+2i Jt \cos q}}{\tan\frac{1}{2}(q-p)}\, .
$
On the finite lattice this function is well defined because $q\ne p$. However, in the thermodynamic
limit $q\sim p$ and the function will have a singularity. We want to extract the
singular part. Using formula 4.4.7 (1) on page 646 of \cite{PBM}, $\sum_{k=1}^{n-1}\cot(x+k\pi/n)=n \cot (nx)$
we find
\begin{align}\label{iden1}
\sum_q\frac{1}{\tan\frac{1}{2}(q-p)}&=\sum_{j=1}^L\frac{1}{\tan\frac{\pi}{L}(j-k\pm\st/2)}
 =\pm L\cot \left(\frac{\pi\st}{2}\right)\, ,
\end{align}
where the plus (minus) sign corresponds to  $\langle a a^\dagger \rangle$
($\langle a^\dagger a\rangle $). Separating  the singular part in the form
\begin{align*}
e(m,t,p)&=\frac{1}{L}\sum_q\frac{e^{ i m q+2i Jt \cos q}-e^{ i m p+2i Jt \cos p}}{\tan(\frac{q-p}{2})}+
\frac{1}{L}\sum_q\frac{e^{ i m p+2i Jt \cos p}}{\tan(\frac{q-p}{2})}\, ,
\end{align*}
the thermodynamic limit limit of $e(m,t,p)\rightarrow E(m,t,p)$ is
\begin{align}\label{TLe}
E(m,t,p)
&=\frac{1}{2\pi}\intp dq\,  \frac{e^{ i m q+2i Jt \cos q}-e^{ i m p+2i Jt \cos p}}{\tan\frac{1}{2}(q-p)}
\pm \cot\left(\frac{\pi\st}{2}\right)e^{ i m p+2i Jt \cos p}\, ,\nonumber\\
&=\mbox{PV }\frac{1}{2\pi}\intp dq\,  \frac{e^{ i m q+2i Jt \cos q}}{\tan\frac{1}{2}(q-p)}
\pm \cot\left(\frac{\pi\st}{2}\right)e^{ i m p+2i Jt \cos p}
\end{align}
where in the last line we have used the identity $\mbox{PV}\frac{1}{2\pi}\intp dq/\tan\frac{1}{2}(q-p)=0$
and the plus (minus) sign corresponding to  $\langle a a^\dagger \rangle$ ($\langle a^\dagger a\rangle $).

The $d(m,t,p,\st)$ function can be written as
\begin{align*}
d(m,t,p,\st)&=\frac{\sin^2(\pi\st/2)}{L^2}\left[\sum_q\frac{e^{ i m q+2i Jt \cos q}-e^{ i m p+2i Jt \cos p}}{\sin^2\frac{1}{2}(q-p)}+
\sum_q\frac{e^{ i m p+2i Jt \cos p}}{\sin^2\frac{1}{2}(q-p)}\right]\, .
\end{align*}
The sum appearing last term in the square parenthesis can be computed using the
formula 4.4.6 (9) on page 645 of \cite{PBM}, $\sum_{k=0}^{n-1}\sin^{-2}(x+k\pi/n)=n^2\sin^{-2}(nx)$,
with the result
\begin{align}\label{iden2}
\sum_q\frac{1}{\sin^2\frac{1}{2}(q-p)}&=\sum_{j=1}^L\frac{1}{\sin^2(\frac{\pi}{L}(j-k\pm\st/2))}
 =\frac{L^2}{\sin^2\left(\frac{\pi\st}{2}\right)}\, ,
\end{align}
Introducing
$
f(p)=\sum_q\frac{e^{ i m q+2i Jt \cos q}-e^{ i m p+2i Jt \cos p}}{\tan\frac{1}{2}(q-p)}
$
and $l(p)=e^{ i m p+2i Jt \cos p}$ it can be shown  that
\[
d(m,t,p,\st)=\frac{2\sin^2(\pi\st/2)}{L^2}\left[\frac {df(k)}{dk}+\frac{dl(k)}{dk}\sum_q\frac{1}{\tan\frac{1}{2}(q-p)}\right]+l(k)\, ,
\]
which together with (\ref{iden1}) and (\ref{TLe}) shows that in the thermodynamic limit $d(m,t,p,\st)\rightarrow D(m,t,p,\st)$
is given by
\be\label{TLd}
D(m,t,p,\st)=e^{ i m p+2i Jt \cos p}+\frac{2\sin^2(\pi\st/2)}{L}\frac{\6}{\6k}E(m,t,p)\, .
\ee

\end{document}